\documentclass[prd,preprintnumbers,amsmath,amssymb,floatfix]{revtex4}
\usepackage{amsmath,amsfonts,latexsym,amssymb,graphicx,graphics,epsfig,subfigure,color,makeidx,cases}
\usepackage{xcolor,diagbox}
\usepackage{multirow}
\usepackage[colorlinks,linkcolor=blue,anchorcolor=blue,citecolor=green,urlcolor=blue]{hyperref}
\usepackage{mathrsfs}

\begin{document}
\title{Quasinormal Ringing of de Sitter Braneworlds}

\author{Hai-Long Jia$^{a}$$^{b}$}
\email{jiahl2024@lzu.edu.cn}
\author{Wen-Di Guo$^{a}$$^{b}$}
\email{guowd@lzu.edu.cn}
\author{Yu-Xiao Liu$^{a}$$^{b}$\footnote{liuyx@lzu.edu.cn, corresponding author}}
\author{Qin Tan$^{c}$}
\email{tanqin@hunnu.edu.cn}

\affiliation{
$^{a}$Lanzhou Center for Theoretical Physics, 
    Key Laboratory for Quantum Theory and Applications of the Ministry of Education,
    Key Laboratory of Theoretical Physics of Gansu Province,
    School of Physical Science and Technology,
    Lanzhou University, Lanzhou 730000, China \\
$^{b}$Institute of Theoretical Physics \& Research Center of Gravitation,
    Lanzhou University, Lanzhou 730000, China \\
$^{c}$Department of Physics,
    Key Laboratory of Low Dimensional Quantum Structures and Quantum Control of Ministry of Education,
    Synergetic Innovation Center for Quantum Effects and Applications,
    Hunan Normal University,
    Changsha, 410081, Hunan, China
}

\begin{abstract}
    Compared with the Poincaré braneworld, the de Sitter (dS) braneworld aligns more closely with 
    the present universe characterized by a small but finite cosmological constant.
    To explore the quasinormal ringing properties within the dS brane scenario, 
    we investigate the gravitational perturbations in both thin and thick dS brane configurations. 
    Analysis of the perturbation equations reveals that 
    the effective potential along the extra dimension exhibits the shape of Pöschl-Teller potential, 
    asymptotically approaching a constant value (mass gap) at infinity. 
    In particular, this mass gap can be reduced. 
    And analytical calculations further indicate that 
    the gravitational perturbations, apart from the zero mode, possess a series of discrete, purely imaginary quasinormal modes in the late stages. 
    This result implies that these perturbations decay without oscillation over time.
    The analytical findings also demonstrate that the brane structure 
    primarily determines the distribution of the quasinormal spectrum 
    while preserving the purely imaginary nature of the quasinormal frequencies. 
    Subsequently, we further simulate the gravitational wave signal by numerically evolving the perturbation equations,
    which yield late-stage results consistent with the analytical predictions.
    Interestingly, these quasinormal modes carry information about the cosmological constant on the brane, 
    which provides a potential new pathway for the study of cosmology in the dS brane scenario. 
\end{abstract}

\maketitle
% \tableofcontents 
%%%%%%%%%%%%%%%%%%%%%%%%%%%%%%%%%%%%%%%%%%%%%%%%%%%%%%%%%%%%%%%%%%%%%%%%%%%%%%%%%%%%%%%%%%%%%%%%%%%%%%%%%%%%%%%%%%%%%%%%%%%%%
\section{Introduction} \label{Introduction}

At the end of the twentieth century, 
the emergence of the braneworld concept~\cite{Polchinski:1995mt} rekindled interest in extra dimensions
and provided a new perspective for solving the hierarchy problem between the electroweak scale and Planck scale~\cite{Antoniadis:1990ew,Arkani-Hamed:1998jmv,Antoniadis:1998ig,Randall:1999ee}.
In this concept, to avoid conflicts with existing particle physics experiments which have not yet detected signals of extra dimensions, 
it is generally required that the fields of the standard model are confined to a four-dimensional hypersurface, commonly referred to as a brane,
while gravity can propagate throughout the entire bulk spacetime.
Based on this concept, various models attempting to solve the hierarchy problem have been proposed, 
such as the compactified large extra dimension model proposed by Arkani-Hamed \textit{et al}.~\cite{Arkani-Hamed:1998jmv,Antoniadis:1998ig}, 
and the warped extra dimension model proposed by Randall and Sundrum (RS)~\cite{Randall:1999ee}.
However, in the RS model, the structure of the brane along the extra dimensions is neglected, 
i.e., the energy density of the brane is a delta function along the extra dimension. 
Therefore, this type of model is referred to as the thin braneworld (3-brane) model. 
In order to eliminate this singularity, DeWolfe \textit{et al}. introduced the concept of domain walls~\cite{Akama:1982jy,Rubakov:1983bb} 
into the infinitely large extra dimension model~\cite{Randall:1999vf}, 
and proposed the so-called thick braneworld (domain wall).     
That is, by replacing the cosmological constant in the bulk spacetime with a bulk dynamical scalar field,
one can endow the internal structure of the brane~\cite{DeWolfe:1999cp,Gremm:1999pj,Csaki:2000fc}.  
This makes the braneworld theory more comprehensive. 
In the subsequent two decades, various braneworld models~\cite{Kaloper:1998sw,Kaloper:1999sm,Nihei:1999mt,Kogan:1999wc,Karch:2000ct,Gregory:2000jc,Dvali:2000rv,Wang:2002pka,Herrera-Aguilar:2010ehj,Gremm:2000dj,Dzhunushaliev:2010fqo,Guo:2011wr,Liu:2012gv,Barbosa-Cendejas:2017vgm,Nakas:2023yhj} 
emerged to address the residual issues of extra dimensions, 
such as the size of extra dimensions~\cite{Rubakov:2001kp,Lee:2020zjt,Branchina:2023rgi}, 
the fine-tuning problem between the bulk cosmological constant and the brane tension~\cite{Karch:2000ct}, 
the reproduction of Newtonian gravity~\cite{Callin:2004py,Guo:2010az,Araujo:2011fm}, 
the localization of matter fields~\cite{Liu:2009ve,Liu:2009uca}
and so on. 
Numerous solutions based on these models have been proposed to solve these problems~\cite{Csaki:2000fc,Csaki:2000pp,Brevik:2002yj,Melfo:2006hh,Gabadadze:2006jm,Liu:2011wi,Xie:2015dva,Zhong:2016iko,Zhou:2017xaq,Xie:2021ayr,Zhu:2023tzx,Zhu:2024gvl}.
For more information on braneworld models, one can refer to Refs.~\cite{Dzhunushaliev:2009va,Maartens:2010ar,Liu:2017gcn,Ahluwalia:2022ttu}.

It is known that branes can have various symmetries. 
The RS models with one and two branes satisfying four-dimensional Poincaré symmetry were considered in 1999~\cite{Randall:1999ee,Randall:1999vf}. 
Subsequently, bent branes were investigated in Refs.~\cite{DeWolfe:1999cp,Kaloper:1999sm,Nihei:1999mt,Karch:2000ct},
where the 3-branes or domain walls satisfy the four-dimensional maximal spacetime symmetries: de Sitter (dS) or anti-de Sitter (AdS). 
The corresponding metrics can be uniformly expressed 
in the following form:
\begin{equation}
    \label{line-element-source}
    ds^2 = e^{2A(y)} ds_4^2  + dy^2, 
\end{equation}
where $ds_4^2$ is labeled as the following four-dimensional metrics that satisfy the maximal symmetry of the four-dimensional spacetime:
\begin{equation}
    ds_4^2 = \left\{
    \begin{aligned}
        &-dt^{2}+dx_{1}^{2}+dx_{2}^{2}+dx_{3}^{2},  &\quad&\text{Poincaré brane,}\\
        &-dt^{2}+e^{2 \alpha t}\left(dx_{1}^{2}+dx_{2}^{2}+dx_{3}^{2}\right),  &\quad&\text{dS brane,}\\
        &e^{2 \tilde{\alpha}  x_3}\left(-dt^{2}+dx_{1}^{2}+dx_{2}^{2}\right)+dx_{3}^{2},  &\quad&\text{AdS brane.}
    \end{aligned}
    \right.
\end{equation}
Here, $\alpha$ and $\tilde{\alpha}$ are related to the effective cosmological constant of the brane. 
For instance, in special cases, $\Lambda_4= 3\alpha^2$ and $\Lambda_4=-3\tilde{\alpha}^2$ for the dS and AdS branes, respectively. 
Moreover, the exploration of bent dS branes can better correspond to the present universe. 
It is generally believed that dS braneworld slices correspond to the flat Friedmann-Robertson-Walker universe, 
while AdS branes play a crucial role in the AdS/CFT correspondence~\cite{Maldacena:1997re}.
Additionally, in the RS model, achieving a zero cosmological constant on the brane requires fine-tuning between the bulk cosmological constant and the brane tension. 
However, it is worth noting that, compared to the Poincaré brane solution, 
the fine-tuning issue becomes significantly more relaxed in the dS and AdS brane solutions when both the bulk and boundary cosmological constants are present.

After establishing a braneworld model, studying gravitational perturbations within the brane scenario is of significant importance.
For the flat brane in the RS model, 
there exists a graviton zero mode and a continuous spectrum of massive Kaluza-Klein (KK) modes for gravitational perturbations~\cite{Randall:1999vf}, 
which introduce corrections to the four-dimensional Newtonian potential at a small scale. 
Subsequently, Csaki \textit{et al}. explored gravitational resonances in flat braneworlds~\cite{Csaki:2000fc,Csaki:2000pp}. 
In 2005, Seahra further investigated the quasinormal modes (QNMs) in the RS model and revealed a potential connection between these two types of modes~\cite{Seahra:2005wk,Seahra:2005iq,Clarkson:2005mg}. 
Some of us have discovered a connection between resonant states and QNMs by studying gravitational perturbations in thick flat branes~\cite{Tan:2023cra}. 
Further calculations and analyses of QNMs in flat branes were conducted in Refs.~\cite{Tan:2022vfe,Tan:2024url,Jia:2024pdk,Tan:2024aym,Tan:2024qij}.
For AdS branes, Karch \textit{et al}. discovered that 
there is a series of discrete KK mass spectra in gravitational perturbations, 
and the corresponding wavefunctions are generally normalizable~\cite{Karch:2000ct,Liu:2009uca,Liu:2011zy,Afonso:2006gi}. 
From this perspective, the AdS brane does not possess QNMs but rather a series of normal modes, 
which play a crucial role in the corrections to the four-dimensional Newtonian potential and in the observation of massive gravitons.
However, for dS branes, some previous articles~\cite{Kaloper:1999sm,Nihei:1999mt,Brevik:2002yj,Araujo:2011fm,Liu:2009dt,Guo:2011qt,Zhong:2023eeq} revealed that 
the gravitational perturbations on these branes may exhibit a finite discrete KK spectrum along with a series of continuous spectrum. 
A notable characteristic is the presence of a mass gap between the starting point of the continuous spectrum and the zero mode. 
This makes it feasible to search for QNMs within these continuous spectra. 
Understanding these modes is crucial for grasping how gravity propagates in the dS brane and 
how it influences the background gravitational wave signals in the present universe. 
Therefore, discussing the QNMs in dS braneworlds is of significant importance.

In this paper, capital Latin letters $M,N,\dots = 0,1,2,3,5$, 
Greek letters $\mu,\nu,\dots=0,1,2,3$, and Latin letters $i,j,\dots=1,2,3$, respectively,
are used to denote the coordinates of the five-dimensional bulk spacetime, four-dimensional spacetime, and three-dimensional space. 
The organization of the paper is as follows: 
In Sec.~\ref{Review of dS brane}, we mainly review the concept of the dS brane scenario, 
providing a brief review of thin and thick branes, and referencing classical results from previous researches. 
Subsequently, we derive the wave equations governing the transverse-traceless tensor (gravitational) perturbations and the Schrödinger-like equations.  
In Sec.~\ref{QNMs of dS brane}, 
we first discuss the properties of the gravitational effective potential.
Next, by using analytical method and continued fraction method, 
we calculate the QNMs of the gravitational perturbations at the late time in various dS branes. 
At last, we simulate the gravitational wave signal by numerically evolving the wave equation, 
analyzing the properties of the signal and discussing potential observational effects.
Finally, we provide a conclusion in Sec.~\ref{conclusion}. 

%%%%%%%%%%%%%%%%%%%%%%%%%%%%%%%%%%%%%%%%%%%%%%%%%%%%%%%%%%%%%%%%%%%%%%%%%%%%%%%%%%%%%%%%%%%%%%%%%%%%%%%%%%%%%%%%%%%%%%%%%%%%%
\section{Review of de Sitter braneworld} \label{Review of dS brane}

In this paper, we consider a dS 3-brane (or domain wall) with the metric ansatz 
\begin{equation}
    \label{line-element}
    ds^2 = e^{2A(z)}\left(\gamma_{\mu\nu}dx^{\mu}dx^{\nu}+dz^2\right), 
\end{equation}
where $A(z)$ is the warp factor and $\gamma_{\mu\nu}=\mathrm{diag}\left\{-1,e^{2\alpha t},e^{2\alpha t},e^{2\alpha t}\right\}$ is the four-dimensional dS metric, 
i.e., the spatially flat Friedmann-Robertson-Walker metric. Here, $\alpha$ is referred to as the Hubble constant and characterizes the expansion rate of the universe on the brane. 
And $z$ denotes the conformal flat coordinate of the extra dimension coordinate $y$ given in Eq.~\eqref{line-element-source}. 
Thus, according to the metric~\eqref{line-element}, we can derive the five-dimensional Einstein tensor as follows:
\begin{align}
    \label{G-munu}
    G_{\mu\nu} &= \gamma_{\mu\nu}\left(3A'^2+ 3A''-3\alpha^2 \right), \\
    \label{G-5mu}
    G_{5\mu} &= 0, \\
    \label{G-55}
    G_{55} &= 6A'^2-6\alpha^2, 
\end{align}
where the prime denotes the derivative with respect to the coordinate $z$. 

%%%%%%%%%%%%%%%%%%%%%%%%%%%%%%%%%%%%%%%%%%%%%%%%%%%%%%%%%%%%
\subsection{The thin dS brane model}

Let us review the model describing a thin dS brane embedded in a five-dimensional bulk spacetime with a cosmological constant. 
The action is~\cite{Randall:1999ee,Randall:1999vf,Gregory:2000jc,Gabadadze:2006jm}
\begin{equation}
    \label{thin-action}
    S = \int d^4x\int_{z>0}dz\,\sqrt{-g}\left(2\hat{M}^3 R - \Lambda_+\right)
    + \int d^4x\int_{z<0}dz\,\sqrt{-g}\left(2\hat{M}^3 R - \Lambda_-\right)
    - \int d^4x\,\sqrt{-g^{\mathrm{brane}}}\,\sigma ,
\end{equation}
where $\hat{M}$ is the five-dimensional Planck mass scale, 
$\sigma$ is the brane tension,
and $g^{\mathrm{brane}}_{\mu\nu}$ is the induced metric on the brane located at $z=0$, 
i.e., the purely four-dimensional components of the bulk metric $g^{\mathrm{brane}}_{\mu\nu}=g_{\mu\nu}(x^{\lambda},z=0)$.  
Here, we set $4\hat{M}^3=1$ for convenience. 

By varying the action~\eqref{thin-action} with respect to the metric,
we can obtain the field equation                                                                      
\begin{equation}
    \label{thin-field-equation}
    \sqrt{-g}\left(R_{MN}-\frac{1}{2}g_{MN}R\right) = 
    -  \left[\Lambda_5 \sqrt{-g}\,g_{MN} 
    + \sigma \sqrt{-g^{\mathrm{brane}}}\,\delta^{\mu}_{M}\delta^{\nu}_{N}g^{\mathrm{brane}}_{\mu\nu}\delta(z)\right] ,
\end{equation}
with the bulk cosmological constant
\begin{equation}
     \Lambda_5 =  \epsilon(z) \Lambda_+  + \epsilon(-z) \Lambda_-,  
\end{equation}
where $\epsilon(z)$ is a step function, i.e., $\epsilon(z)$ is $1$ for $z>0$, $\tfrac{1}{2}$ for $z=0$, and $0$ for $z<0$. 
Furthermore, substituting Eqs.~\eqref{G-munu} and~\eqref{G-55} to Eq.~\eqref{thin-field-equation}, 
we can get 
\begin{align}
    \label{thin-field-equation-munu}
    (\mu,\nu):&& \gamma_{\mu\nu}\left(3A'^2+ 3A''-3\alpha^2 \right) &= -e^{2A} \gamma_{\mu\nu}\left( \Lambda_5 + e^{-A} \sigma \delta(z)\right),\\ 
    \label{thin-field-equation-55}
    (5,5):&& 6A'^2-6\alpha^2 &=  - e^{2A} \Lambda_5. 
\end{align}

Note that for a flat brane, the $\alpha^2$ terms in the above equations vanish.
On the other hand, due to the presence of the thin brane,  
we require the warp factor to be continuous at the location of the brane,   %continuous across the position of the brane, 
with the jump in its first derivative determined by the brane tension. 
Specifically, by integrating both sides of Eq.~\eqref{thin-field-equation-munu} over the small interval $\left(-\epsilon, +\epsilon \right)$, 
we can derive the jump condition of $A(z)$:
\begin{equation}
    \label{junction-conditions}
    A'\Big|_{-\epsilon}^{+\epsilon}=-\frac{1}{3} e^{A(z=0)} \sigma  ,
\end{equation}

%%%%%%%%%%%%%%%%%%%%%%%%%%%%%%%%%%%%%%%%%%%%%%%%%%%%%%%%%%%%%%%%%%%%%%%%%%%%%%%%%%%
\subsubsection{Solutions for \texorpdfstring{$\Lambda_+ = \Lambda_-$}{}}\label{Lambda5-Symmetric}

First, we consider solutions satisfying the $Z_2$ symmetry, which results in $\Lambda_+ = \Lambda_-$. 
For a thin dS brane embedded in a five-dimensional AdS spacetime, 
the metric is given by~\cite{Kaloper:1999sm,Nihei:1999mt,DeWolfe:1999cp,Karch:2000ct}:
\begin{equation}
    \label{thin-solution-y-AdS}
    A(y)= \ln \left\{\frac{\alpha}{k} \sinh\left[k (y_H-|y|)\right]\right\} , 
\end{equation}
where $y$ is the extra dimension coordinate and $y_H$ represents the position of the cosmological horizon. 
And the bulk cosmological constant $\Lambda_5$ is $\Lambda_5=-6k^2$.
Interestingly, it can be shown that,
by applying the normalization condition, such that $A(0)=0$, 
we can solve the Hubble constant $\alpha$ in terms of 
the cosmological constant $\Lambda_5$ and the brane tension $\sigma$,
\begin{equation}
    \label{relation-L-H-S}
    \alpha^2=\frac{6\Lambda_5+\sigma^2}{36}. 
\end{equation}
It can be observed that, 
unlike the flat brane solution which requires fine-tuning $\Lambda_5=-\tfrac{\sigma^2}{6}$, 
the cosmological constant satisfies an inequality $-\tfrac{\sigma^2}{6} < \Lambda_5 < 0$. 
In addition, when $\Lambda_5<-\tfrac{\sigma^2}{6}$, 
the effective four-dimensional cosmological constant is negative, 
indicating that the brane geometry is AdS$_4$.

\begin{figure*}[htb]
    \begin{center}
    \includegraphics[width=5.6cm]{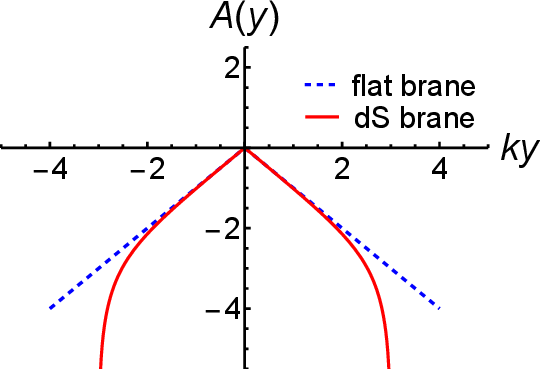}
    \end{center}
    \caption{The warp factors of the flat thin brane and dS one for $y_H=3$ in the AdS bulk spacetime.}
    \label{Figure-1}
\end{figure*}

Using the flat brane solution $A(y)=-k|y|$ of the RS-II model~\cite{Randall:1999vf} as a contrast, 
we plot the behavior of the solution~\eqref{thin-solution-y-AdS} in Fig.~\ref{Figure-1}.
As shown in Fig.~\ref{Figure-1}, within the same AdS bulk spacetime, 
the addition of a boundary cosmological constant on the brane causes $A(y)$ to diverge at $y= \pm y_H$, 
indicating two coordinate singularities (cosmological horizons). 
However, it can be demonstrated that gravity is still localized on the brane in the usual way.
In the coordinate $z$, the position of the horizon is stretched to infinity.

\begin{figure*}[htb]
    \begin{center}
    \subfigure[~AdS and flat bulk]  {\label{AdS-thin-brane-z}
    \includegraphics[width=5.6cm]{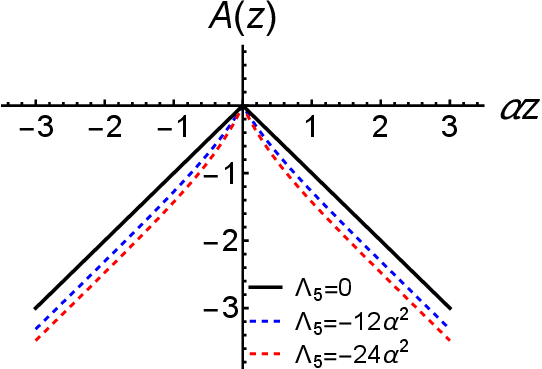}}
    \subfigure[~dS and flat bulk]  {\label{dS-thin-brane-z}
    \includegraphics[width=5.6cm]{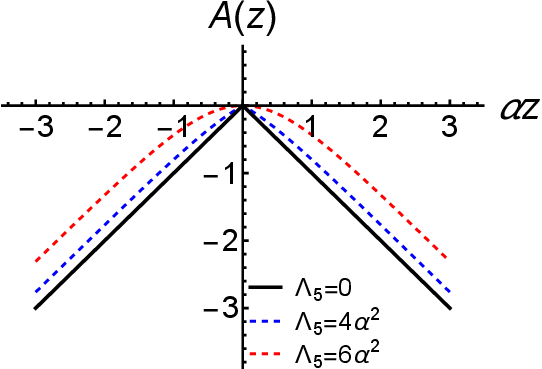}}
    \end{center}
    \caption{The shapes of the warp factors $A(z)$ of the thin dS brane in the AdS, dS, and flat bulk spacetimes.}
    \label{Figure-2}
\end{figure*}

Furthermore, the AdS bulk spacetime can be extended to Minkowski or dS spacetimes. 
This yields a series of distinct solution families for thin dS brane solutions in $z$ coordinate. 
According to the relation~\eqref{relation-L-H-S} and the values of $\Lambda_5$, 
the classifications are shown as below~\cite{Cvetic:1999ec}:
\begin{itemize}
    \item When $\Lambda_{5} =-6k^2 < 0$, the thin dS brane is embedded in an AdS$_5$ spacetime, the warp factor is 
    \begin{equation}
        \label{thin-dS-solution-AdS}
        A(z)=-\ln \left[\frac{k}{\alpha} \sinh\left(\alpha (|z|+z_{0})\right) \right] ,
    \end{equation}
    where $z_0=\mathrm{arcsinh} (\alpha/k)/\alpha$, and the brane tension is $\sigma = 6\alpha \coth(\alpha z_{0})$. 
    This solution is nothing more than the result of the solution~\eqref{thin-solution-y-AdS} in the $z$ coordinate. 
    \item When $\Lambda_{5} =0$, the entire spacetime has zero curvature, resulting in the dS brane solution mentioned in Ref.~\cite{Guerrero:2002ki},
    \begin{equation}
        \label{thin-dS-solution-flat}
        A(z)= -\alpha |z|.
    \end{equation}
    The brane tension is $\sigma = 6 \alpha$. 
    \item When $0 <\Lambda_{5}=6k^2 \leq 6\alpha^2$, 
    the bulk spacetime is dS$_5$, with the dS$_4$ brane embedded as a codimension one hypersurface. 
    Specifically, in the limit $\Lambda_{5} \rightarrow  6\alpha^2$, the brane tension vanishes, 
    and the thin brane extends into a thick brane, resulting in a thick dS brane structure, as detailed in Sec.~\ref{Model-D}.
    The warp factor can be expressed as
    \begin{equation}
        \label{thin-dS-solution-dS}
        A(z)=-\ln \left[\frac{k}{\alpha} \cosh\left(\alpha (|z|+z_{1})\right) \right] , 
    \end{equation}
    where $z_1=\mathrm{arccosh} (\alpha/k)/\alpha$ and the brane tension is $\sigma = 6\alpha \tanh(\alpha z_{1})$.  
\end{itemize}
The shapes of the warp factors $A(z)$ associated with the three types of geometry are shown in Fig.~\ref{Figure-2}.

%%%%%%%%%%%%%%%%%%%%%%%%%%%%%%%%%%%%%%%%%%%%%%%%%%%%%%%%%%%%%%%%%%%%%%%%%%%%%%%%%%%
\subsubsection{Solutions for \texorpdfstring{$\Lambda_+ \neq \Lambda_-$}{}}
\label{Model-B}

Next, we consider the asymmetric warped geometry, that is $\Lambda_+ \neq \Lambda_-$. 
In 2006, Gabadadze \textit{et al}. studied the impact of an asymmetric warped geometry on gravitational perturbations in a flat brane and
discovered that when the AdS curvature scales on both sides of the brane differ, a resonance mode appears~\cite{Gabadadze:2006jm}. 
This resonance mode was found to have a complementary relation with the RS zero mode. 
Thereafter, in 2011, Araujo \textit{et al}. investigated the dS brane scenario and found a resonance mode in the gravitational perturbation spectrum~\cite{Araujo:2011fm}. 
As the cosmological constant $\Lambda_5$ decreases, the contribution of this resonance mode to the Newtonian potential increases. 
The corresponding form of the warp factor is
\begin{equation}
    \label{thin-solution-unify}
    A(z)=\left\{
    \begin{aligned}
        &-\ln \left[ e^{\alpha z} - K_+ \sinh \alpha z\right], &&z\geq 0, \\
        &-\ln \left[ e^{-\alpha z} + K_- \sinh \alpha z\right], &&z<0, 
    \end{aligned}
    \right.
\end{equation}
where 
\begin{equation}
    \label{K-formula}
    K_{\pm} = 1-\sqrt{1-\frac{\Lambda_{\pm}}{6\alpha^2}}
\end{equation}
and
the brane tension is
\begin{equation}
    \label{relation-tension}
    \sigma = 3\alpha(2 - K_+ - K_-) , \quad \Lambda_{\pm} \leq 6\alpha^2. 
\end{equation}
From Eq.~\eqref{relation-tension}, it is evident that 
in order to ensure the brane tension is a real number, 
the five-dimensional cosmological constant must satisfy the inequality $\Lambda_{\pm} \leq 6\alpha^2$. 
This corresponds exactly to the case discussed in Sec.~\ref{Lambda5-Symmetric}. 
For $0 <\Lambda_{\pm}=6k_{\pm}^2 \leq 6\alpha^2$, 
the bulk spacetime is dS$_5$, and the relationship between the two solutions is given by 
$K_{\pm} =1- k_{\pm} \sinh(\alpha z_1)/\alpha $.
For $\Lambda_{\pm}=0$, 
the bulk spacetime is flat, and $K_{\pm}=0$.
For $\Lambda_{\pm}=-6k_{\pm}^2<0$,
the bulk spacetime is AdS$_5$, and $K_{\pm} = 1-k_{\pm} \cosh(\alpha z_0)/\alpha $.

On the one hand, based on Eqs.~\eqref{thin-solution-unify} and~\eqref{K-formula}, 
the asymmetric solutions provide a unified description for the three bulk geometries.
Additionally, in the scenario where $\Lambda_{\pm}=0$, 
these solutions can seamlessly transition between different cases. 
However, the symmetric solutions do not exhibit this favorable behavior. 
For example, in the limit $\Lambda_{5} \rightarrow 0$,
the solution~\eqref{thin-dS-solution-AdS} leads to $z_0 \rightarrow \infty$,
causing the solution to diverge, and it cannot smoothly revert to the solution~\eqref{thin-dS-solution-flat}. 
And a similar issue arises with the solution~\eqref{thin-dS-solution-dS}.
On the other hand, the correspondence between the symmetric and asymmetric solutions 
shows that the asymmetric geometric construction can be viewed as a blockwise (or segmented) combination of symmetric geometries, 
which helps to simplify the subsequent analysis.

%%%%%%%%%%%%%%%%%%%%%%%%%%%%%%%%%%%%%%%%%%%%%%%%%%%%%%%%%%%%
\subsection{A thick dS brane generated by a bulk scalar field}

Now, we review a thick dS braneworld generated by a bulk scalar field. 
The action is expressed as 
\begin{equation}
    \label{thick-action}
    S = \int d^4x dz \sqrt{-g} \left( 2\hat{M}^3 R - \frac{1}{2}\nabla_{M}\phi \nabla^{M}\phi - V(\phi) \right), 
\end{equation}
where $\hat{M}^3$ is also set to $1/4$, $\phi$ represents the bulk scalar field, and $V(\phi)$ is the bulk scalar potential. 
We assume $\phi=\phi(z)$. 
The five-dimensional Einstein's equation and the bulk scalar field equation are   
\begin{align}
    \label{thick-field-equation}
    R_{MN}-\frac{1}{2}g_{MN}R &= g_{MN} \left( - \frac{1}{2}\nabla_{P}\phi \nabla^{P}\phi - V(\phi) \right) + \nabla_{M}\phi \nabla_{N}\phi ,\\
    \label{phi-field-equation}
    g^{MN}\nabla_{M}\nabla_{N}\phi &= \frac{\partial V(\phi)}{\partial \phi} . 
\end{align}
Furthermore, substituting Eqs.~\eqref{G-munu} and~\eqref{G-55} to Eqs.~\eqref{thick-field-equation} and~\eqref{phi-field-equation}, 
we can obtain 
\begin{align}
    \label{thick-field-equation-munu}
    (\mu,\nu)&:&  \gamma_{\mu\nu}\left(3A'^2+ 3A''-3\alpha^2 \right) &= \gamma_{\mu\nu}\left[-\frac{1}{2}\phi'^2-e^{2A} V(\phi)\right] ,\\
    \label{thick-field-equation-55}
    (5,5)&:&  6A'^2-6\alpha^2 &= \frac{1}{2}\phi'^2-e^{2A} V(\phi) ,\\
    \label{phi-field-equation-1}
    \mathrm{scalar}&:& \phi''+3 A' \phi' &= e^{2A} \frac{\partial V(\phi)}{\partial \phi} .  
\end{align}
Through calculation, we can see that Eq.~\eqref{phi-field-equation-1} can be derived from Eqs.~\eqref{thick-field-equation-munu} and~\eqref{thick-field-equation-55}. 
That is, the above equations are not independent of each other. 
Therefore, we must specify one of the functions $A(z)$, $\phi(z)$, and $V(\phi)$ to solve these equations, 
or alternatively, one can use the superpotential method~\cite{DeWolfe:1999cp,Bazeia:2008zx}.

%%%%%%%%%%%%%%%%%%%%%%%%%%%%%%%%%%%%%%%%%%%%%%%%%%%%%%%%%%%%
\subsubsection{Solutions for \texorpdfstring{$\phi \neq 0$}{}}
\label{Model-C}

In the thick braneworld generated by a scalar field, 
Wang has proposed a class of analytical solutions with the metric ansatz~\eqref{line-element} in the $z$ coordinate~\cite{Wang:2002pka}, as follows:

\begin{align}
    \label{thick-solution-phineq0-Az}
    A(z) &=-n\ln \left[\cosh\left(\beta z\right)\right], \\
    \label{thick-solution-phineq0-phi}
    \phi(z) &=\phi_{0}\sin^{-1}\left[\tanh\left(\beta z\right)\right],\\ 
    \label{thick-solution-phineq0-Vphi}
    V(\phi) &=V_{0} \cos^{2(1-n)}\left(\frac{\phi}{\phi_{0}}\right), 
\end{align}
where $n$ and $\beta$ are arbitrary constants with the relation $\alpha^2=n^2 \beta^2$, and $\phi_0 = \sqrt{3n(1-n)}$, $V_0 = n\beta^2[3(1+3n)]/2$. 
The range of $n$ is restricted to $0<n<1$. 

\begin{figure*}[htb]
    \begin{center}
    \subfigure[~Warp factor]  {\label{Az-Plot}
    \includegraphics[width=5.6cm]{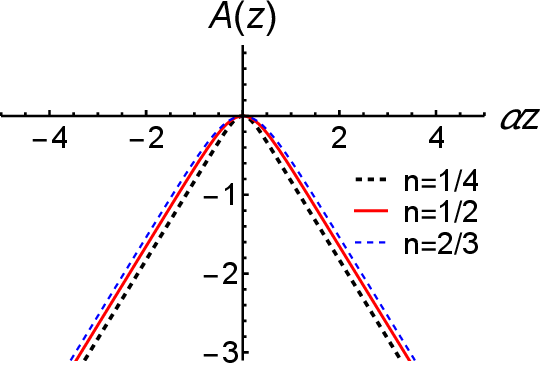}}
    \subfigure[~Bulk scalar field]  {\label{phiz-Plot}
    \includegraphics[width=5.6cm]{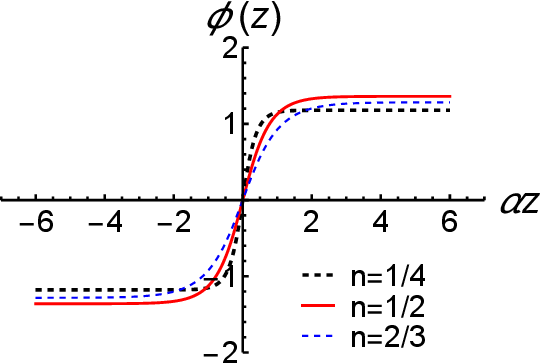}}
    \subfigure[~Scalar potential]  {\label{Vphiz-Plot}
    \includegraphics[width=5.6cm]{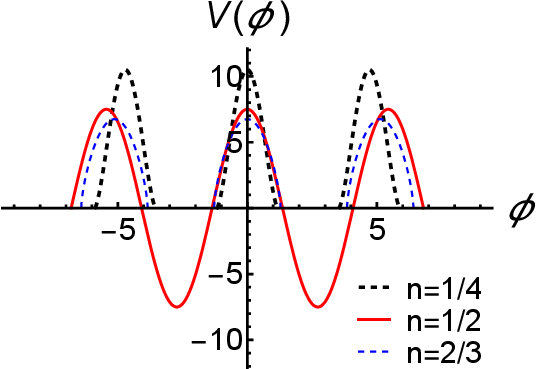}}
    \end{center}
    \caption{The shapes of the warp factor $A(z)$, the bulk scalar field $\phi(z)$, and the self-interaction scalar potential $V(\phi)$.}
    \label{Figure-3}
\end{figure*}

Plots of the warp factor~\eqref{thick-solution-phineq0-Az}, 
the bulk scalar field~\eqref{thick-solution-phineq0-phi}, 
and the bulk scalar potential~\eqref{thick-solution-phineq0-Vphi} are shown in Fig.~\ref{Figure-3}. 

\begin{figure*}[htb]
    \begin{center}
    \includegraphics[width=5.6cm]{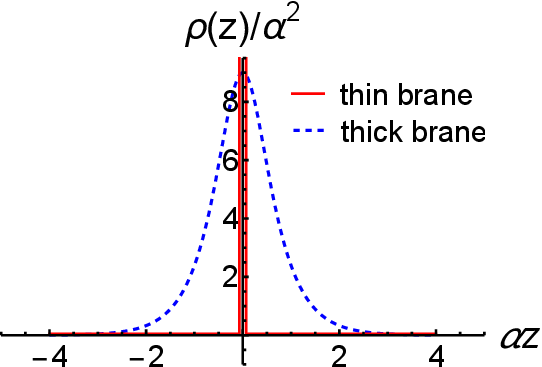}
    \end{center}
    \caption{The distinction of the energy densities of the thin brane and the thick brane for $n=\tfrac{1}{2}$.}
    \label{Figure-4}
\end{figure*}

Comparing Figs.~\ref{Figure-2} and~\ref{Az-Plot}, we observe that 
the main distinction between the warp factors of thin and thick branes 
lies in the continuity of their first derivatives at $z=0$. 
As shown in Eq.~\eqref{junction-conditions}, the first derivative of the warp factor is associated with the brane tension in the case of a thin brane. 
Conversely, the warp factor in a thick brane is smooth throughout, 
as the brane extends beyond a simple hypersurface to exhibit a smooth distribution along the extra dimension. 
This difference is further illustrated by the energy densities of the branes,
\begin{align}
    \text{thin brane:}& &&\rho(z) = \sigma\delta(z),     \\
    \text{thick brane:}& &&\rho(z) = \frac{1}{2}e^{-2A}\phi'^2 + V(\phi) = \frac{3n(1+n)\beta^2}{\cosh^{2(1-n)}\left(\beta z\right)},       
\end{align}
as depicted in Fig.~\ref{Figure-4}.

%%%%%%%%%%%%%%%%%%%%%%%%%%%%%%%%%%%%%%%%%%%%%%%%%%%%%%%%%%%%
\subsubsection{Solutions for \texorpdfstring{$\phi = 0$}{}}
\label{Model-D}

In the above solution, the value of $n$ has an upper limit. 
It can be seen that when $n=1$, the scalar field $\phi$ vanishes 
and the bulk scalar potential $V(\phi)$ becomes a constant. 
In this case, the scalar potential can be equivalently treated as the bulk positive cosmological constant, 
and the action~\eqref{thick-action} can be rewritten as 
\begin{equation}
    \label{thick-action-re}
    S = \int d^4x dz \sqrt{-g} \left( 2\hat{M}^3 R - \Lambda_5\right) . 
\end{equation}
The solution of the warp factor is~\cite{Herrera-Aguilar:2010ehj}
\begin{equation}
    \label{thick-solution-phieq0}
    A(z) =-\ln \left[\frac{b}{\alpha} \cosh\left(\alpha z\right)\right], 
\end{equation}
where $6b^2= \Lambda_5$. The introduction of $b$ is to distinguish between the bulk cosmological constant 
and the four-dimensional effective cosmological constant $\Lambda_4 = 3 \alpha^2$.

Applying the normalization condition, we can obtain $b=\alpha$.
Thus, we obtain the same warp factor solution as presented in Ref.~\cite{Gremm:2000dj}, 
despite it focusing on a flat brane scenario.
And the solution is consistent with the limit case $\Lambda_5 \rightarrow 6\alpha^2$ of the solution~\eqref{thin-dS-solution-dS}.
The gravitational perturbations along the extra dimension, specifically the KK mass spectrum, exhibit the same behavior.
This consistency arises because, in a braneworld structure governed by the metric~\eqref{line-element-source}, 
this behavior is determined exclusively by the warp factor.
Additionally, the existence of this solution also represents that a special case of obtaining the thick brane without using a scalar field, 
which is a pure dS braneworld. 

%%%%%%%%%%%%%%%%%%%%%%%%%%%%%%%%%%%%%%%%%%%%%%%%%%%%%%%%%%%%
\subsection{Tensor perturbation} \label{Tensor perturbation}

\begin{figure*}[htb]
    \begin{center}
    \subfigure[~thin dS brane]  {\label{psi0AB}
    \includegraphics[width=5.6cm]{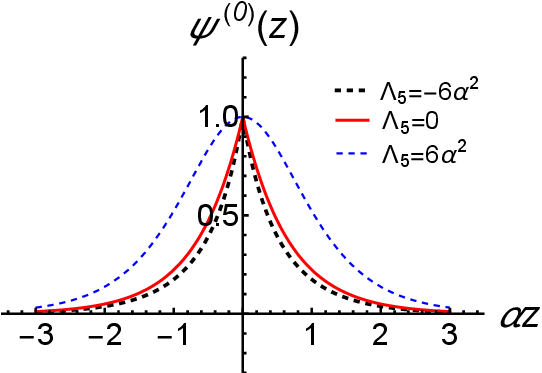}}
    \subfigure[~thick dS brane]  {\label{psi0CD}
    \includegraphics[width=5.6cm]{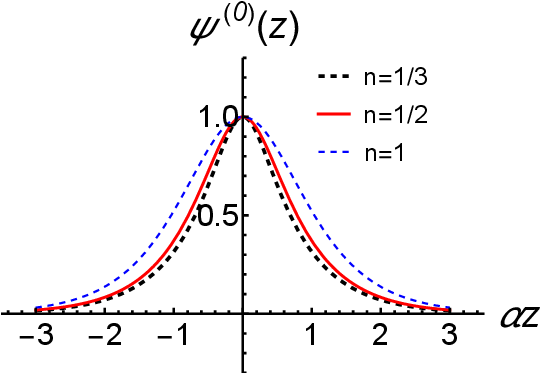}}
    \end{center}
    \caption{The shapes of the zero mode $\psi^{(0)}(z)$ for the tensor perturbation on the dS brane. }
    \label{Figure-add-1}
\end{figure*}

In this subsection, we investigate the tensor perturbation in the braneworld models. 
First, we consider the perturbed metric as the following form 
\begin{equation}
    \label{line-element-perturbation}
    ds^2=e^{2A(z)}\left[\left(\gamma_{\mu\nu}+h_{\mu\nu}\right)dx^{\mu}dx^{\nu}+dz^2\right] , 
\end{equation}
where $h_{\mu\nu}(x^{\lambda},z)$ satisfies the transverse-traceless condition $\gamma^{\mu\nu}h_{\mu\nu}=0=\gamma^{\mu\nu}\nabla_{\mu}h_{\nu\lambda}$. 
The gauge introduced here is referred to as the RS gauge, which is appropriate for the five-dimensional braneworld context.
By substituting the perturbed metric~\eqref{line-element-perturbation} 
into Eq.~\eqref{thin-field-equation} or~\eqref{thick-field-equation}, 
we can get the perturbed equation satisfied by $h_{\mu\nu}$,
\begin{equation}
    \label{five-wave-equation}
    \left[\partial_{z}^{2} + 3 \left(\partial_{z} A(z)\right)\partial_{z} + \square^{(4)}-2\alpha^2\right] h_{\mu\nu} = 0,
\end{equation}
where $\square^{(4)}=\gamma^{\alpha\beta}\nabla_{\alpha}\nabla_{\beta}$ is the four-dimensional D'Alembert operator. 
And our calculations indicate that, regardless of the brane structure, 
the perturbation $h_{\mu\nu}$ always satisfies Eq.~\eqref{five-wave-equation}.
The influence of the brane structure is primarily embedded in the terms related to the extra dimensional components.
Then we make the ansatz $ h_{\mu\nu}= e^{-\frac{3}{2}A(z)} X_{\mu\nu}(x^{\lambda}) \psi(z)$. 
Equation~\eqref{five-wave-equation} can be decomposed as two parts: 
\begin{align}
    \label{four-wave-equation}
    \left(\square^{(4)}-2\alpha^2\right) X_{\mu\nu} &= m^2 X_{\mu\nu} , \\
    \label{extra-dimensional-part}
    \left( -\partial_z^2 + V(z) \right) \psi(z) &= m^2 \psi(z),  
\end{align}
where
\begin{equation}
    \label{effective-potential}
    V(z) = \frac{3}{2}A''(z) + \frac{9}{4} A'^2(z)
\end{equation}
is the effective potential of the tensor perturbation. 
Thus, it can be seen that the equation for the extra dimensional part, Eq.~\eqref{extra-dimensional-part}, is a Schrödinger-like equation. 
By choosing appropriate boundary conditions, one can obtain the KK mass spectrum. 
In particular, the zero mode that can be localized on the brane is always present.
And it satisfies $\psi^{(0)}(z) \propto e^{\frac{3}{2}A(z)}$, as shown in Fig.~\ref{Figure-add-1}.
Meanwhile, the equation for the four-dimensional components, Eq.~\eqref{four-wave-equation}, 
describes the propagation of gravitational waves on the brane. 
Furthermore, by combining Eqs.~\eqref{four-wave-equation} and~\eqref{extra-dimensional-part}, 
we can study the problem of gravitational localization on the dS braneworld. 
Some previous articles~\cite{Kaloper:1999sm,Nihei:1999mt,Brevik:2002yj,Wang:2002pka,Liu:2009dt,Guo:2010az,Araujo:2011fm} have suggested that 
the gravitational perturbation on dS branes support a finite discrete KK spectrum along with a series of continuous KK spectrum. 

Furthermore, we can perform scalar-vector-tensor decomposition for $h_{\mu\nu}$ with respect to the spatial metric $\gamma_{ij}$.  
We only study the tensor part, which describes the gravitational waves on the brane.
Thus, in this situation, the perturbed metric is
\begin{equation}
    \label{line-element-perturbation-hij}
    ds^2=e^{2A(z)}\left[-dt^2+ e^{2\alpha t}\left(\delta_{ij}+h_{ij}\right)dx^{i}dx^{j}+dz^2\right] , 
\end{equation}
where $h_{ij}(x^{\lambda},z)$ satisfies $\delta^{ij}h_{ij}=0=\delta^{ij}\partial_{i}h_{jk}$.
By substituting the metric~\eqref{line-element-perturbation-hij} into Eq.~\eqref{thin-field-equation} or~\eqref{thick-field-equation},
we can obtain the perturbed equation of $h_{ij}$,
\begin{equation}
    \label{five-wave-equation-hij}
    \left[-\left(\partial_{t}^{2}+3\alpha \partial_{t}\right) + e^{-2\alpha t} \partial_{k}\partial^{k} + \partial_{z}^{2} + 3 \left(\partial_{z} A(z)\right)\partial_{z} \right] h_{ij} = 0.
\end{equation}
For a spatially flat universe, Eq.~\eqref{five-wave-equation-hij} is equivalent to $\square^{(5)}h_{ij}=0$. 
The $x^i$-dependence of $h_{ij}$ can be separated from the $t$- and $z$-dependence. 
Thus, we can assume 
\begin{equation}
    \label{hij-assumption}
    h_{ij}=\varepsilon_{ij} e^{-ip_{k}x^{k}} \Phi(t,z), 
\end{equation}
where
\begin{equation}
    \label{Phi-assumption}
    \Phi(t,z) = e^{-\frac{3}{2}\alpha t} e^{-\frac{3}{2}A(z)} \Psi(t,z). 
\end{equation}
Substituting it into Eq.~\eqref{five-wave-equation-hij}, we have
\begin{equation}
    \label{psi-equation}
    \left[-\partial_{t}^{2} + \frac{9}{4} \alpha ^2 - e^{-2 \alpha  t} p^2  + \partial_{z}^{2}  -V(z) \right] \Psi (t,z) =0, 
\end{equation}
where $p^2=\delta^{ij}p_{i}p_{j}$. 
Unlike the equation for the gravitational perturbation $h_{ij}$ on a flat brane, 
Eq.~\eqref{psi-equation} contains the term $9\alpha ^2/4$ and 
the exponential factor $e^{-2 \alpha  t}$ in front of $p^2$. 
The term $9\alpha ^2/4$ precisely cancels out with the mass gap in the effective potential $V(z)$ in Eq.~\eqref{effective-potential}, 
while the exponential factor $e^{-2 \alpha  t} p^2$ prevents us 
from using the ansatz $e^{-i\omega t}$ to separate variables and obtain a Schrödinger-like equation, 
making it challenging to solve Eq.~\eqref{psi-equation} either analytically or semi-analytically.
However, we find that in the late stage of the evolution, assuming the evolution starts from $t_0$, 
the effect of this time-dependent term becomes negligible because of the exponential factor $e^{-2 \alpha  t}$ . 

Therefore, we can consider the case of $p=0$,
which not only allows for an analytical solution to obtain the quasinormal spectrum 
but also corresponds to the late-time behavior of the solution to Eq.~\eqref{psi-equation}. 
Correspondingly, when $p=0$, Eq.~\eqref{psi-equation} becomes to 
\begin{equation}
    \label{psi-equation-late-time}
    \left[-\partial_{t}^{2}  + \partial_{z}^{2}  -V(z) + \frac{9}{4} \alpha ^2 \right] \Psi (t,z) =0.   
\end{equation}
For the case of $p \neq 0$, 
numerical methods can be applied to solve Eq.~\eqref{psi-equation}.

In next section, our discussion will first focus on 
the quasinormal behavior of perturbations for the case $p=0$, 
which also corresponds to the late-time quasinormal behavior for $p \neq 0$. 
This will primarily be achieved by solving Eq.~\eqref{psi-equation-late-time}. 
Subsequently, we can apply numerical evolution methods to solve Eq.~\eqref{psi-equation} 
and obtain the perturbation behavior throughout the entire evolution process.

%%%%%%%%%%%%%%%%%%%%%%%%%%%%%%%%%%%%%%%%%%%%%%%%%%%%%%%%%%%%%%%%%%%%%%%%%%%%%%%%%%%%%%%%%%%%%%%%%%%%%%%%%%%%%%%%%%%%%%%%%%%%%
\section{Quasinormal ringing of de Sitter branes} \label{QNMs of dS brane}

In the framework of the braneworld scenario, 
gravitational waves are permitted to propagate along the extra-dimensional direction. 
By applying an analogous method to black hole perturbation analysis, 
one can extract critical information about the brane 
through the study of gravitational perturbations. 
In a manner similar to how QNMs of black hole perturbations encode essential physical properties of the black hole, 
the QNMs of a brane provide insight into the intrinsic characteristics of the brane. 
This allows the investigation of brane properties through the analysis of its perturbation spectrum.

As mentioned at the end of Sec.~\ref{Tensor perturbation}, 
the presence of the term $e^{-2 \alpha  t} p^2$ makes the $e^{-i\omega t}$ assumption inapplicable.
However, in the late-stage evolution of Eq.~\eqref{psi-equation}, this term becomes negligible. 
Therefore, the late-time behavior of the solution to Eq.~\eqref{psi-equation} 
can be more conveniently obtained by solving Eq.~\eqref{psi-equation-late-time}, 
specifically the $p=0$ case of Eq.~\eqref{psi-equation}. 
In this case, we can apply the assumption $\Psi (t,z) \sim e^{-i\omega t} \psi(z) $ 
to Eq.~\eqref{psi-equation-late-time},
reducing it to 
\begin{equation}
    \label{extra-dimensional-equation-re}
    \left( -\partial_z^2 + V_{\text{re}}(z) \right) \psi(z) = \omega^2 \psi(z),  
\end{equation}
where $V_{\text{re}}(z)= V(z) - 9\alpha ^2 / 4$ is the reduced effective potential. 
Comparing this equation with Eq.~\eqref{extra-dimensional-part}, 
we also derive the following relation: 
\begin{equation}
    \omega^{2} = m^2 - \frac{9}{4} \alpha ^2.
\end{equation}
Note that this relation holds only in the case of $p=0$
or in the late stages of gravitational perturbations when $p \neq 0$.

In this section, we will calculate the QNMs of dS branes. 
Before proceeding with the calculations, 
we first explore the nature of the effective potential of gravitational perturbations,
since the solutions to Schrödinger-like equation~\eqref{extra-dimensional-equation-re} are primarily determined by the behavior of the effective potential.
Then, we solve Eq.~\eqref{extra-dimensional-equation-re} analytically to obtain the QNMs, 
and verify the results using the continued fraction method.
Finally, we perform a numerical evolution of Eq.~\eqref{psi-equation} 
and discuss the features of these signals and their potential impact on gravitational wave observations.

%%%%%%%%%%%%%%%%%%%%%%%%%%%%%%%%%%%%%%%%%%%%%%%%%%%%%%%%%%%%%%%%%%%%%%%%%%%%%%%%%%%%%%%%%%%%%%%%%%%%%%%%%%%%%%%%%%%%%%%%%%%%%
\subsection{Properties of effective potentials} \label{Properties-of-effective-potentials}

\begin{figure*}[htb]
    \begin{center}
    \subfigure[~$\Lambda_5 \leq  0 $]{\label{VzthinAdS}
    \includegraphics[width=5.6cm]{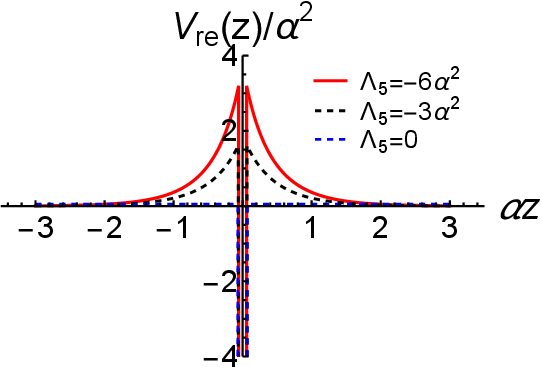}}
    \subfigure[~$0 < \Lambda_5 \leq 6\alpha^2 $]{\label{VzthindS}
    \includegraphics[width=5.6cm]{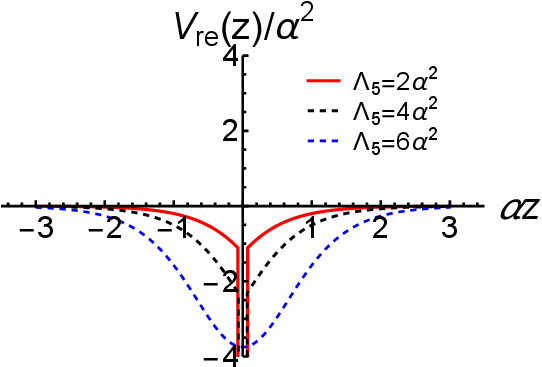}}
    % \vskip\floatsep
    \subfigure[~$\Lambda_- < 0 < \Lambda_+ < 6\alpha^2 $]{\label{VzthindSAdS}
    \includegraphics[width=5.6cm]{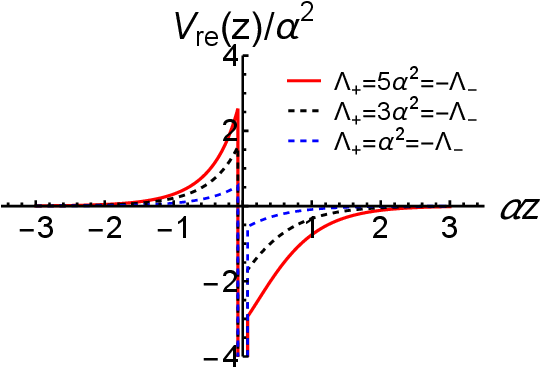}}
    \end{center}
    \caption{The shapes of the reduced effective potentials of the thin dS brane for the different bulk spacetimes.}
    \label{Figure-5}
\end{figure*}

In this subsection, we discuss the characteristics of the (reduced) effective potential. 
Starting with the thin brane case, 
we can derive the specific expressions of the reduced effective potentials 
by combining Eqs.~\eqref{thin-dS-solution-AdS},~\eqref{thin-dS-solution-flat},~\eqref{thin-dS-solution-dS}, 
and~\eqref{thin-solution-unify} with Eq.~\eqref{effective-potential}, as follows:
\begin{equation}
    \label{effective-potential-thin-symmetry}
    \text{Symmetry: } \qquad  V_{\text{re}}(z)=\left\{
    \begin{aligned}
        & \frac{15\alpha^2}{4}\frac{1}{\sinh^2 \left(\alpha (|z|+z_{0})\right) } - 3\alpha\; \coth(\alpha z_{0}) \delta(z) , &&\text{for AdS bulk},  \\
        & -3\alpha\;\delta(z) , &&\text{for flat bulk},  \\
        & - \frac{15\alpha^2}{4}\frac{1}{\cosh^2 \left(\alpha (|z|+z_{1})\right) } - 3\alpha\; \tanh(\alpha z_{1}) \delta(z) ,  &&\text{for dS bulk} , 
    \end{aligned}
    \right.
\end{equation}
\begin{equation}
    \label{effective-potential-thin-antisymmetry}
    \text{Asymmetry: } \qquad  V_{\text{re}}(z)= 
    - 15\alpha^2 \left[\left(\frac{\sqrt{\frac{6\alpha^2}{\Lambda_+}} K_+}{ e^{\alpha z}  + \frac{6\alpha^2}{\Lambda_+} K_+^2  e^{-\alpha z} } \right)^2 \epsilon(z) + \left(\frac{\sqrt{\frac{6\alpha^2}{\Lambda_-}} K_-}{ e^{-\alpha z}  + \frac{6\alpha^2}{\Lambda_-} K_-^2  e^{\alpha z} } \right)^2 \epsilon(-z)\right]
    - \frac{\sigma}{2} \delta(z). 
\end{equation}
The shapes of the reduced effective potentials of the thin dS brane are shown in Fig.~\ref{Figure-5}.

From Fig.~\ref{Figure-5}, we can see that, 
as categorized at the end of Sec.~\ref{Lambda5-Symmetric}, 
different properties of the bulk spacetime result in distinct shapes of the effective potential.
Notably, there is a delta function in the effective potential at $z=0$,
which requires applying an Israel junction condition~\cite{Israel:1966rt} to the perturbation $h_{ij}$ 
at the location of the brane, specifically
\begin{equation}
    \label{Israel-junction-condition}
    \left[ \partial_{z} \psi\right]_{z=0} = - \sigma \psi(z=0),
\end{equation}
where $\left[ \partial_{z} \psi\right]_{z=0} = \partial_{z} \psi |_{z=0^+} - \partial_{z} \psi |_{z=0^-}$. 
If the brane satisfies $Z_2$ symmetry, we have $\partial_{z} \psi |_{z=0^+} = - \partial_{z} \psi |_{z=0^-}$, 
and the junction condition~\eqref{Israel-junction-condition} becomes 
\begin{equation}
    \label{Israel-junction-condition-symmetry}
    \partial_{z} \psi |_{z=0^+} = - \frac{\sigma}{2} \psi(z=0). 
\end{equation}
Particularly, when $\Lambda_5 = 6\alpha^2$, 
the delta function in the effective potential disappears, 
corresponding to the transition from a thin brane to a thick brane configuration.

\begin{figure*}[htb]
    \begin{center}
    \subfigure[~$\phi \neq 0$]  {\label{Vzthickn}
    \includegraphics[width=5.6cm]{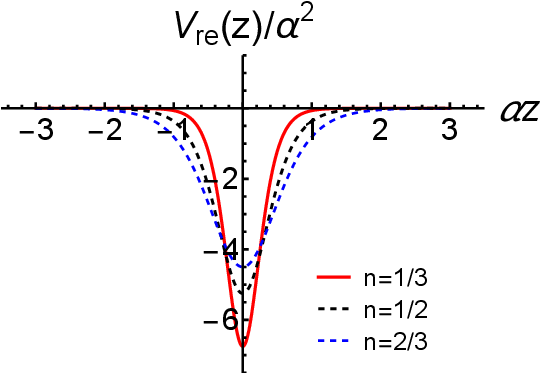}}
    \subfigure[~$\phi = 0$]  {\label{Vzthickn1}
    \includegraphics[width=5.6cm]{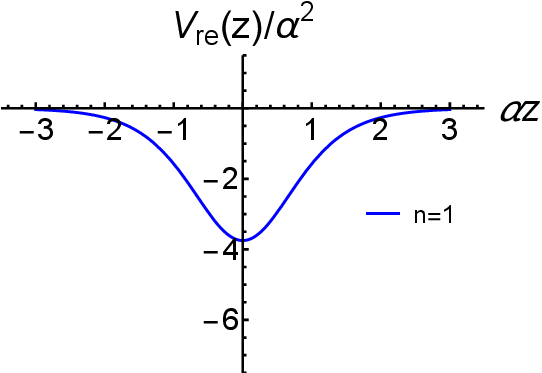}}
    \end{center}
    \caption{The shapes of the reduced effective potentials~\eqref{effective-potential-thick-phineq0} and~\eqref{effective-potential-thick-phieq0} of the thick dS brane.}
    \label{Figure-6}
\end{figure*}

Next, we discuss the thick brane scenario. 
Similarly, by substituting Eqs.~\eqref{thick-solution-phineq0-Az} and~\eqref{thick-solution-phieq0} 
into Eq.~\eqref{effective-potential}, 
we obtain the explicit form of the reduced effective potential as follows:
\begin{align}
    \label{effective-potential-thick-phineq0}
    &\phi \neq 0 \; (0<n<1) : \quad  &&V_{\text{re}}(z)= - \frac{3n(3n+2)\alpha^2}{4 n^2} \frac{1}{\cosh^2(\alpha z/n)} , \\
    \label{effective-potential-thick-phieq0}
    &\phi = 0 \; (n=1) : \quad     &&V_{\text{re}}(z)= - \frac{15\alpha^2}{4} \frac{1}{\cosh^2(\alpha z)} .
\end{align}
The shapes of the effective potentials on the thick dS brane are shown in Fig.~\ref{Figure-6}.

In particular, Fig.~\ref{Vzthickn1} aligns with the blue dashed line in Fig.~\ref{VzthindS}, 
as expected, since the warp factors of these two cases are identical. 

Note that, the reduced effective potential 
differs from the effective potential by the mass gap term $9\alpha^2/4$. 
Additionally, comparing Figs.~\ref{Figure-2} and~\ref{Az-Plot} with Figs.~\ref{Figure-5} and~\ref{Figure-6}, 
we can find the key distinctions between the thin and thick brane cases: 
(1) the thin brane exhibits a discontinuity in the first derivative of the warp factor, 
and (2) the effective potential features a delta function configration in the thin brane case, 
which is absent in the thick brane case.
On the other hand, 
from Eqs.~\eqref{effective-potential-thin-symmetry},~\eqref{effective-potential-thin-antisymmetry},
\eqref{effective-potential-thick-phineq0}, and~\eqref{effective-potential-thick-phieq0}, 
it can be observed that the reduced effective potential takes the similar form with a Pöschl-Teller (PT) potential. 
In the black hole perturbation theory, 
the PT potential is often used as an approximation to the true effective potential, 
providing approximate values for black hole QNMs. 
Interestingly, in braneworld scenarios, 
PT potentials are commonly observed, 
making the treatment of gravitational perturbations on the brane analytically solvable. 
The study of PT potentials in quantum mechanics has a long history, 
with detailed discussions available in Refs.~\cite{rosen1932vibrations,poschl1933bemerkungen,Witten:1981nf,D_az_1999}.

%%%%%%%%%%%%%%%%%%%%%%%%%%%%%%%%%%%%%%%%%%%%%%%%%%%%%%%%%%%%%%%%%%%%%%%%%%%%%%%%%%%%%%%%%%%%%%%%%%%%%%%%%%%%%%%%%%%%%%%%%%%%%
\subsection{Quasinormal modes of dS branes} \label{Quasinormal modes of dS branes}

Before solving Eq.~\eqref{extra-dimensional-equation-re}, 
we need to clearly specify the boundary conditions. 
For the QNM problem in such a brane system,
we follow the approach used in the black hole QNM analysis, 
choosing the outgoing boundary conditions, that is, only outgoing waves should exist at infinity.

By examining Eq.~\eqref{psi-equation-late-time}, 
we find that the last two terms on the left-hand side vanish as $z \rightarrow \pm \infty$, 
reducing Eq.~\eqref{psi-equation-late-time} to 
\begin{equation}
    \label{wave-equation-ztoinfinity}
    \left[-\partial_{t}^{2} + \partial_{z}^{2} \right] \Psi(t,z)=0 .
\end{equation}
The general solution is $\Psi(t,z) \propto e^{-i \omega (t\pm z)}$. 
With the assumption $\Psi (t,z) \sim e^{-i\omega t} \psi(z) $,
the specific outgoing boundary conditions can be written as 
\begin{equation}
    \label{boundary-condition}
    \psi(z) \propto \left\{
    \begin{aligned}
        & e^{i \omega z} ,  &&z\rightarrow \infty, \\
        & e^{-i \omega z} ,  &&z\rightarrow -\infty. 
    \end{aligned}
    \right.
\end{equation}
Additionally, for the thin branes, 
the Israel junction condition~\eqref{Israel-junction-condition} is required at the brane location, 
whereas this condition is not needed for thick branes.

Now we proceed to solve the Schrödinger-like equation~\eqref{extra-dimensional-equation-re} 
with the boundary conditions~\eqref{boundary-condition} and the Israel junction condition~\eqref{Israel-junction-condition}.
To facilitate the calculations, we nondimensionalize some parameters as follows: 
\begin{equation}
    \label{nondimensionalize}
    \bar{t}=\alpha t, \quad \bar{z}=\alpha z, \quad \bar{\omega} = \omega/\alpha.
\end{equation}

%%%%%%%%%%%%%%%%%%%%%%%%%%%%%%%%%%%%%%%%%%%%%%%%%%%%%%%%%%%%%%%%%%%%%%%%%%%%%%%%%%%%%%%%%%%%%%%%%%%%%%%%%%%%%%%%%%%%%%%%%%%%%
\subsubsection{Thin branes} \label{Quasinormal modes of dS thin branes}

Since the asymmetric solution includes all cases covered by the symmetric solution, 
we only need to focus on discussing the asymmetric solution~\eqref{thin-solution-unify} in detail. 
Furthermore, the potential functions on either side of the brane are separated by a delta function
but share a similar form. 
Thus, we can handle the right and left regions separately, 
and then connect the results from both sides using the Israel junction condition.

Now, we start to solve the QNMs of the gravitational perturbations of the thin dS brane. 
By substituting Eq.~\eqref{effective-potential-thin-antisymmetry} 
into Eq.~\eqref{extra-dimensional-equation-re}, 
we can obtain the equations describing the left and right regions of the brane, 
given by the following formulas:

\begin{align}
    \label{extra-dimensional-equation-re-right}
    &\left[ - \partial_{\bar{z}}^{2} - 15 \left(\frac{\sqrt{\frac{6\alpha^2}{\Lambda_+}} K_+}{ e^{\bar{z}}  + \frac{6\alpha^2}{\Lambda_+} K_+^2  e^{-\bar{z}} } \right)^2 \right] \psi_+(\bar{z})= \bar{\omega}^2 \psi_+(\bar{z}) , \quad \bar{z}>0, \\
    \label{extra-dimensional-equation-re-left}
    &\left[ - \partial_{\bar{z}}^{2} - 15 \left(\frac{\sqrt{\frac{6\alpha^2}{\Lambda_-}} K_-}{ e^{-\bar{z}}  + \frac{6\alpha^2}{\Lambda_-} K_-^2  e^{\bar{z}} } \right)^2 \right] \psi_-(\bar{z})= \bar{\omega}^2 \psi_-(\bar{z}) , \quad \bar{z}<0. 
\end{align}

Next, we perform two coordinate transformations 
$\hat{z}_+=\frac{1}{ 1  + \frac{6\alpha^2}{\Lambda_+} K_+^2  e^{-2 \bar{z}} }$ and $\hat{z}_-=\frac{1}{ 1  + \frac{6\alpha^2}{\Lambda_-} K_-^2  e^{2 \bar{z}} }$, 
then the boundary conditions~\eqref{boundary-condition} are rewritten as
\begin{equation}
    \label{boundary-condition-re}
    \left\{
    \begin{aligned}
        &\psi_+(\hat{z}_+) \propto (1-\hat{z}_+)^{-i \bar{\omega}/2} ,  && \hat{z}_+\rightarrow 1 \; (z \rightarrow \infty), \\
        &\psi_-(\hat{z}_-) \propto (1-\hat{z}_-)^{-i \bar{\omega}/2} ,  && \hat{z}_-\rightarrow 1 \; (z \rightarrow -\infty).
    \end{aligned}
    \right.
\end{equation}
We set $\psi_{\pm}(\hat{z}_{\pm}) \propto \hat{z}_{\pm}^{-i \bar{\omega}/2} \left(1-\hat{z}_{\pm}\right)^{-i \bar{\omega}/2} \xi_{\pm}(\hat{z}_{\pm})$, 
where $\hat{z}_{\pm}^{-i \bar{\omega}/2}$ serves as an auxiliary component to facilitate the solution. 
Substituting it into Eqs.~\eqref{extra-dimensional-equation-re-right} and~\eqref{extra-dimensional-equation-re-left}, 
we can obtain two standard hypergeometric equations for $\xi_{+}(\hat{z}_{+})$ and $\xi_{-}(\hat{z}_{-})$: 
\begin{align}
    \label{PT-equation-re-HS-right}
    \hat{z}_{+}\left(1-\hat{z}_{+}\right) \frac{d^2 \xi_{+}}{d \hat{z}_{+}^2} + \left[h_3-(h_1+h_2+1)\hat{z}_{+}\right]\frac{d \xi_{+}}{d \hat{z}_{+}} -h_1 h_2 \xi_{+} =0, \\
    \label{PT-equation-re-HS-left}
    \hat{z}_{-}\left(1-\hat{z}_{-}\right) \frac{d^2 \xi_{-}}{d \hat{z}_{-}^2} + \left[h_3-(h_1+h_2+1)\hat{z}_{-}\right]\frac{d \xi_{-}}{d \hat{z}_{-}} -h_1 h_2 \xi_{-} =0, 
\end{align}
where $h_1=-i\bar{\omega}-\frac{3}{2}$, $h_2=-i\bar{\omega}+\frac{5}{2}$, and $h_3=1-i\bar{\omega}$. 
The coefficients $h_1$, $h_2$, and $h_3$ of the hypergeometric equations on both sides of the brane are found to be consistent.
The general solutions of Eqs.~\eqref{PT-equation-re-HS-right} and~\eqref{PT-equation-re-HS-left} are 
$\xi_{+}(\hat{z}_{+}) = D_{_{+}}\cdot {}_2\!F_1 \left(h_1,h_2,h_3,\hat{z}_{+}\right)$ 
and $\xi_{-}(\hat{z}_{-}) = D_{_{-}}\cdot {}_2\!F_1 \left(h_1,h_2,h_3,\hat{z}_{-}\right)$, respectively.
Thus, the solutions of Eqs.~\eqref{extra-dimensional-equation-re-right} and~\eqref{extra-dimensional-equation-re-left} are 
\begin{align}
    \label{PT-equation-sol-QNMs-right}
    \psi_+(\hat{z}_{+}) =  D_{+} \cdot \left(\hat{z}_+(1-\hat{z}_+)\right)^{-i \bar{\omega}/2} {}_2\!F_1 \left(h_1,h_2,h_3,\hat{z}_+\right) , \\
    \label{PT-equation-sol-QNMs-left}
    \psi_-(\hat{z}_{-}) =  D_{-} \cdot \left(\hat{z}_-(1-\hat{z}_-)\right)^{-i \bar{\omega}/2} {}_2\!F_1 \left(h_1,h_2,h_3,\hat{z}_-\right) . 
\end{align}
When $z \rightarrow \pm\infty$ ($\hat{z}_{\pm}\rightarrow 1$), $\psi_{\pm}(\hat{z}_{\pm}\rightarrow 1)\thicksim \left(1-\hat{z}_{\pm}\right)^{-i \bar{\omega}/2}$, using 
\begin{align}
    \label{F-formula}
    {}_2\!F_1 \left(h_1,h_2,h_3,\hat{z}\right) =& (1-\hat{z})^{i \bar{\omega}} \frac{\Gamma (h_3) \Gamma(h_1+h_2-h_3)}{\Gamma(h_1)\Gamma(h_2)} {}_2\!F_1 \left(h_3-h_1,h_3-h_2,h_3-h_1-h_2+1,1-\hat{z}\right) \nonumber \\
    +& \frac{\Gamma (h_3) \Gamma(h_3-h_1-h_2)}{\Gamma(h_3-h_1)\Gamma(h_3-h_2)} {}_2\!F_1 \left(h_1,h_2,-h_3+h_1+h_2+1,1-\hat{z}\right),
\end{align}
we find that the boundary conditions~\eqref{boundary-condition-re} 
require that the first term on the right-hand side of Eq.~\eqref{F-formula} vanishes, 
meaning $\tfrac{1}{\Gamma(h_1)}=0$ or $\tfrac{1}{\Gamma(h_2)}=0$.
Thus, we can obtain the QNMs that satisfy the following relationship:
\begin{equation}
    \label{QNMs-result-thin}
    \bar{\omega} = \left( \pm 2 - \frac{2 N +1}{2}\right) i , \qquad N=0,1,2,\dots,
\end{equation}
where $N$ is the overtone index. 
Clearly, the results show that all of these frequencies are purely imaginary.

Note that, it can be seen that the QNMs obtained on both sides are consistent. 
For the matching of the two solutions on both sides of the brane, 
we need to apply the Israel junction condition~\eqref{Israel-junction-condition}, 
which yields the following relationship between the coefficients of the solutions on both sides:
\begin{align}
    \frac{D_+}{D_-} &= \frac{\Lambda_+}{\Lambda_-}\left(\frac{\Lambda_+}{\alpha ^2}\right)^{i \bar{\omega} /2} \left(\frac{\Lambda_-}{\alpha ^2}\right)^{-i \bar{\omega} /2} 
    \left[ -96\alpha K_- (3+2i\bar{\omega} )  \; {}_2\!F_1 \left(h_1+1,h_2,h_3,\frac{\Lambda_-}{12 \alpha ^2 K_- }\right) \right. \nonumber \\
    &- \left. \left( 6\Lambda_- \sqrt{1-\frac{\Lambda_+}{6\alpha ^2}} +6\Lambda_+ \sqrt{1-\frac{\Lambda_-}{6\alpha ^2}} +48\alpha\left(6\sqrt{1-\frac{\Lambda_-}{6\alpha ^2}}-4i\bar{\omega}-6\right)\right) \; {}_2\!F_1 \left(h_1,h_2,h_3,\frac{\Lambda_-}{12 \alpha ^2 K_- }\right) \right] \Bigg{/} \nonumber \\
    &  \left[ \left( 6\Lambda_+ \left( \sqrt{1-\frac{\Lambda_+}{6\alpha ^2}}+\sqrt{1-\frac{\Lambda_-}{6\alpha ^2}}\right) - 48\alpha\left(6\sqrt{1-\frac{\Lambda_+}{6\alpha ^2}}-4i\bar{\omega}-6\right) \right) \; {}_2\!F_1 \left(h_1,h_2,h_3,\frac{\Lambda_+}{12 \alpha ^2 K_+ }\right)  \right. \nonumber \\
    &  \left. -96\alpha K_+ (3+2i\bar{\omega} )  \; {}_2\!F_1 \left(h_1+1,h_2,h_3,\frac{\Lambda_+}{12 \alpha ^2 K_+ }\right) \right] .
\end{align}

In light of the above, 
through analytical derivation, we find that the QNMs of gravitational perturbations for the thin dS brane in the AdS and dS bulk spacetimes are purely imaginary. 
In particular, regardless of the properties of the bulk spacetime on either side of the thin dS brane (e.g., differing cosmological constants $\Lambda_5 \neq 0$), 
the result given in Eq.~\eqref{QNMs-result-thin} is unique.
And the properties of the spacetime on either side of the brane 
primarily affect the junction conditions of the wave function at the brane, 
resulting in different values of the effective potentials
and different matching values of the wave function at the junction.

%%%%%%%%%%%%%%%%%%%%%%%%%%%%%%%%%%%%%%%%%%%%%%%%%%%%%%%%%%%%%%%%%%%%%%%%%%%%%%%%%%%%%%%%%%%%%%%%%%%%%%%%%%%%%%%%%%%%%%%%%%%%%
\subsubsection{Thick branes} \label{Quasinormal modes of dS thick branes}

For the thick brane, the junction condition is not required,
and it is sufficient to analyze the solution~\eqref{thick-solution-phineq0-Az} alone.
Similarly, substituting Eq.~\eqref{effective-potential-thick-phineq0} 
into Eq.~\eqref{extra-dimensional-equation-re}, 
we have the Schrödinger-like equation: 
\begin{equation}
    \label{PT-equation-re-thick}
    \left[ - \partial_{\bar{z}}^{2} - \frac{3n(3n+2)}{4n^2} \frac{1}{\cosh^2(\bar{z}/n)} \right] \psi(\bar{z})= \bar{\omega}^2 \psi(\bar{z}) ,  
\end{equation}
with the outgoing boundary conditions~\eqref{boundary-condition}.
Next, performing a coordinate transformation $\hat{z}=\frac{1}{ 1  + e^{-2 \bar{z}/n} }$, 
the boundary conditions~\eqref{boundary-condition} are rewritten as
\begin{equation}
    \label{boundary-condition-re-thick}
    \psi(\hat{z}) \propto \left\{
    \begin{aligned}
        &(1-\hat{z})^{-i \bar{\omega}n/2} ,  && \hat{z}\rightarrow 1, \quad \text{i.e.}~ z \rightarrow \infty, \\
        &\hat{z}^{-i \bar{\omega}n/2} ,  && \hat{z}\rightarrow 0, \quad \text{i.e.}~ z \rightarrow -\infty.
    \end{aligned}
    \right.
\end{equation}
Similarly, we set $\psi(\hat{z})=\hat{z}^{-i \bar{\omega}n/2} \left(1-\hat{z}\right)^{-i \bar{\omega}n/2} \xi(\hat{z})$. 
Substituting it into Eq.~\eqref{PT-equation-re-thick},
we can get a standard Hypergeometric equation for $\xi(\hat{z})$: 
\begin{equation}
    \label{PT-equation-HS-thick}
    \hat{z}\left(1-\hat{z}\right) \frac{d^2 \xi}{d \hat{z}^2} + \left[h_3-(h_1+h_2+1)\hat{z}\right]\frac{d \xi}{d \hat{z}} -h_1 h_2 \xi =0, 
\end{equation}
where $h_1=\left(-i\bar{\omega}-\frac{3}{2}\right)n$, $h_2=1+\left(-i\bar{\omega}+\frac{3}{2}\right)n $, and $h_3=1-i\bar{\omega}n$. 
Following a similar approach to the treatment of the thin brane, 
we can finally derive the solution as 
\begin{equation}
    \label{PT-equation-sol-QNMs-thick}
    \psi(\hat{z}) =D_{1}\cdot \left(\hat{z}(1-\hat{z})\right)^{-i \bar{\omega}n/2} {}_2\!F_1 \left(h_1,h_2,h_3,\hat{z}\right),
\end{equation}
with the corresponding QNMs given by 
\begin{equation}
    \label{QNMs-result-thick}
    \bar{\omega} =  \left( \pm \frac{3n+1}{2n} - \frac{2 N +1}{2 n} \right) i , \qquad N=0,1,2,\dots 
\end{equation}
By comparing Eqs.~\eqref{QNMs-result-thin} and~\eqref{QNMs-result-thick}, 
it can be seen that the QNMs of the thin brane exactly are same as to the QNMs of the thick brane in the case $n=1$.

To verify the analytical results~\eqref{QNMs-result-thin} and~\eqref{QNMs-result-thick}, 
we use the continued fraction method~\cite{Leaver:1985ax,Leaver:1990zz} to recalculate the QNMs of gravitational perturbations of the dS brane.
Namely, we need to solve the Schrödinger-like equation~\eqref{PT-equation-re-thick} (or~\eqref{extra-dimensional-equation-re-right} and~\eqref{extra-dimensional-equation-re-left})
with the outgoing boundary conditions~\eqref{boundary-condition}.

First, by performing a coordinate transformation $\tilde{z}=\tanh \left(\bar{z}/n\right)$, 
Eq.~\eqref{PT-equation-re-thick} and the corresponding boundary conditions~\eqref{boundary-condition} can be rewritten as  
\begin{equation}
    \label{PT-equation-re-thick-re}
    \left\{(1-\tilde{z}^2) \partial_{\tilde{z}}^2-2\tilde{z}\partial_{\tilde{z}} +\left[\frac{n^2 \bar{\omega}^2}{1-\tilde{z}^2} + \frac{3n(3n+2)}{4} \right]\right\}\psi(\tilde{z})=0, 
\end{equation}
and
\begin{equation}
    \label{boundary-condition-re2}
    \psi(\tilde{z}) \propto \left\{
    \begin{aligned}
        &(1-\tilde{z})^{-i \bar{\omega}n/2} ,  && \tilde{z}\rightarrow 1, \quad \;\;\; \text{i.e.}~ \bar{z} \rightarrow \infty, \\
        &(1+\tilde{z})^{-i \bar{\omega}n/2} ,  && \tilde{z}\rightarrow -1, \quad \text{i.e.}~ \bar{z} \rightarrow -\infty, 
    \end{aligned}
    \right.
\end{equation}
respectively. Then, following the continued fraction method~\cite{Leaver:1985ax}, 
the wave function $\psi(\tilde{z})$ that satisfies the boundary conditions~\eqref{boundary-condition-re2} is
\begin{equation}
    \psi(\tilde{z}) = (1-\tilde{z})^{-i \bar{\omega}n/2} (1+\tilde{z})^{-i \bar{\omega}n/2} \sum_{N=0}^{\infty}a_N \tilde{z}^N ,
\end{equation}
where $a_N$ is the expansion coefficient. 
Substituting it into Eq.~\eqref{PT-equation-re-thick-re}, we can obtain a two-term recurrence relation: 
\begin{equation} 
    \label{recurrence-relation}
    \rho_N a_{N+2} + \theta_N a_N =0,\quad N=0,1,2,\cdots, 
\end{equation}
where the recurrence coefficients $\rho_N$ and $\theta_N$ are functions of $N$, $\bar{\omega}$, and the parameter $n$: 
\begin{align}
    \rho_N &= N^2 + 2N +2 , \\
    \theta_N &= \left(-N^2 + \frac{3n(3n+2)}{4} + N (2 i \bar{\omega}n-1) + (i\bar{\omega}n  + \bar{\omega}^2 n^2)\right). 
\end{align}
In fact, the recurrence relation~\eqref{recurrence-relation} can be written into the following matrix form: 
\begin{equation}
    \label{matrix-equation}
    \begin{pmatrix}
        \theta_0 & 0 & \rho_0 & & & \\
         & \theta_1 & 0 & \rho_1 & & \\
         &  & \theta_2 & 0 & \rho_2 & \\
         &  &  & \ddots & \ddots & \ddots
    \end{pmatrix}
    \begin{pmatrix}
        a_0 \\
        a_1 \\
        a_2 \\
        \vdots 
    \end{pmatrix}
    =0. 
\end{equation}
It can be seen that the matrix equation~\eqref{matrix-equation} corresponds to the algebraic equation for QNMs. 
By setting the determinant of its coefficient matrix to zero, 
we can obtain the QNMs of this system. 
Generally, this coefficient matrix is infinite-dimensional, 
so in practice, it must be truncated. 
Fortunately, since this coefficient matrix is an upper triangular matrix, 
we can find the solutions directly
\begin{equation}
    \theta_0=0\;,\; \theta_1=0\;,\; \theta_2=0\;,\; \cdots\;,\; \theta_N=0 \;,\; \cdots,
\end{equation}
namely, 
\begin{equation}
    \label{QNMs-result-thick-CFM}
    \bar{\omega} = 
    \left\{
    \begin{aligned}
        & \left(  - \frac{N}{n} + \frac{3}{2}\right) i ,   \\
        & \left( - \frac{N+1}{n} -\frac{3}{2}  \right) i ,
    \end{aligned}
    \right. 
    \qquad N=0,1,2,\cdots. 
\end{equation}
Finally, we find that this result is entirely consistent with the analytical solution. 
Similarly, we can achieve the same result of the thin dS brane.

\begin{figure*}[htb]
    \begin{center}
    \includegraphics[width=5.6cm]{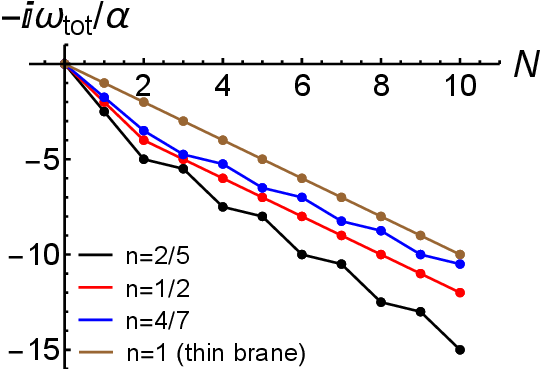}
    \end{center}
    \caption{The first ten QNFs of the dS brane for different values of the parameter $n$.
    The QNMs of the thin dS brane are consistent with those of the thick dS brane when $n=1$.}
    \label{Figure-7}
\end{figure*}

In summary, we can highlight the following points:
\begin{itemize}
    \item The purely imaginary QNF results obtained here 
    characterizes the late-time behavior of gravitational perturbation QNMs.
    This purely imaginary characteristic manifests in the gravitational waveforms, 
    which exhibit pure damping in the late-stage.
    \item These purely imaginary QNFs depend only on the parameter $n$ and the Hubble constant $\alpha$. 
    The parameter $n$ characterizes the internal structure of the brane, 
    and as a result, the brane structure determines the distribution of the quasinormal spectrum. 
    For the thin brane case, the bulk cosmological constant (except for the case of $\Lambda_5 = 0$) 
    has no effect on the quasinormal spectrum.
    \item From the analysis of Eqs.~\eqref{QNMs-result-thin} and \eqref{QNMs-result-thick}, 
    it is evident that the imaginary parts of the first few QNFs are positive.
    However, this does not imply instability, as the consideration is based on the behavior of 
    $\Psi(t,z)$ in Eq.~\eqref{Phi-assumption} rather than $\Phi(t,z)$. 
    As indicated by Eqs.~\eqref{hij-assumption} and~\eqref{Phi-assumption}, 
    the total QNMs of the perturbations $h_{ij}$ should correspond to the solution $\Phi(t,z)$,
    i.e., the factor $e^{-\frac{3}{2}\alpha t}$ multiplied by the solution $\Psi(t,z)$.
    Therefore, the total QNMs of the dS branes are 
    \begin{equation}
        \label{QNMs-result-total}
        \omega^{\text{tot}}=\left(\bar{\omega} -\frac{3i}{2}\right)\alpha = i \left(\pm \frac{3n+1}{2n} - \frac{3n + 1 + 2 N}{2 n}\right) \alpha, \qquad N=0,1,2,\dots ,
    \end{equation}
    as shown in Fig.~\ref{Figure-7}.
    In particular, for both thin and thick branes, the first QNM corresponds to $\omega^{\text{tot}}=0$ ($\bar{\omega}_{0}=3i / 2$), 
    which actually corresponds to the graviton zero mode with $m=0$. 
\end{itemize}
The similar purely imaginary behavior of the QNFs is analogous to 
that of QNMs of gravitational perturbations in odd-dimensional dS spacetimes~\cite{Natario:2004jd}. 
And we have observed the similar results in the gravitational perturbations 
of a Poincaré thick brane with a finite extra dimension~\cite{Jia:2024pdk}. 

%%%%%%%%%%%%%%%%%%%%%%%%%%%%%%%%%%%%%%%%%%%%%%%%%%%%%%%%%%%%%%%%%%%%%%%%%%%%%%%%%%%%%%%%%%%%%%%%%%%%%%%%%%%%%%%%%%%%%%%%%%%%%
\subsection{Time domain evolution of initial wave packet} \label{Time domain evolution}

In the previous parts, we analytically calculated the QNMs for gravitational perturbations for $p=0$,
which also shows the late behavior of gravitational perturbations for $p \neq 0$. 
Now, let us apply numerical evolution methods to solve Eq.~\eqref{psi-equation} 
governing the gravitational perturbations for $p \neq 0$, 
in order to obtain the waveform of the perturbations throughout the entire evolution process. 
Specifically, we consider an initial wave packet evolves under Eq.~\eqref{psi-equation}: 
\begin{equation}
    \label{psi-equation-re}
    \left[-\partial_{t}^{2} + \partial_{z}^{2} - e^{-2 \alpha  t} p^2 -V_{\text{re}}(z) \right] \Psi (t,z) =0. 
\end{equation}

A simple analysis shows that the term $e^{-2 \alpha t} p^2$ does not 
alter the purely imaginary nature of the QNMs. 
Rather, it acts as an overall upward shift in the effective potential, an effect that weakens over time.
Moreover, the lack of a real part in the QNMs makes frequency spectrum analysis inapplicable in this context. 
Additionally, in actual waveform evolution simulations, 
sub-waves with higher decay rates are naturally obscured by those with lower decay rates.
Consequently, it becomes evident that, 
when evolving the wave function based on Eq.~\eqref{psi-equation-re}, 
we can only observe the frequency of the zero mode, $\bar{\omega}_{0}=3i / 2$. 
This finding is later confirmed in our evolution simulations.

To obtain the first QNM from the evolution waveform,
we can utilize the dual form of the effective potential.
As noted in Ref.~\cite{Cooper:1994eh}, the effective potential and its dual potential share the same normal (or quasinormal) spectrum, 
with the dual effective potential lacking precisely the zero mode compared to the effective potential.
This means that the ground state mode of the dual effective potential 
corresponds directly to the first QNM of the effective potential. 
This feature has been confirmed in numerous studies~\cite{Tan:2022vfe,Tan:2023cra,Jia:2024pdk}. 
Therefore, to some extent, we can attempt to propose the ``dual equation'' of Eq.~\eqref{psi-equation-re}, 
\begin{equation}
    \label{psi-equation-re-dual}
    \left[-\partial_{t}^{2} + \partial_{z}^{2} - e^{-2 \alpha  t} p^2 -V_{\text{re}}^{\text{dual}}(z) \right] \hat{\Psi} (t,z) =0, 
\end{equation}
where
\begin{equation}
    V_{\text{re}}^{\text{dual}}(z) = -\frac{3}{2}A''(z) + \frac{9}{4} A'^2(z) - \frac{9\alpha ^2}{4}
\end{equation}
is the reduced dual effective potential. 
Strictly speaking, 
it remains to be proved whether Eqs.~\eqref{psi-equation-re} and~\eqref{psi-equation-re-dual} 
are still strictly dual, due to the existence of the term $e^{-2 \alpha t} p^2$. 
For the case $p=0$, 
it can be shown that a dual relation exists between the two equations.
This duality only ensures that they share the same quasinormal sperctrum, 
but their wavefunctions are different.
However, the reason we evolve the ``dual wavefunction'' is 
to extract the first QNM which is covered by the zero mode 
in the actual waveform. 

Now we can perform the numerical evolution of an initial wave packet 
for Eqs.~\eqref{psi-equation-re} and~\eqref{psi-equation-re-dual}. 
For simplicity, a Gaussian wave packet is chosen as the initial data, 
although other types of wave packets yield the same results.

%%%%%%%%%%%%%%%%%%%%%%%%%%%%%%%%%%%%%%%%%%%%%%%%%%%%%%%%%%%%%%%%%%%%%%%%%%%%%%%%%%%%%%%%%%%%%%%%%%%%%%%%%%%%%%%%%%%%%%%%%%%%%
\subsubsection{Thin branes} \label{Time domain evolution of dS thin branes}

\begin{figure*}[htb]
    \begin{center}
    \subfigure[~$\Phi(t,z_{ext})$]  {\label{thinPhiz1}
    \includegraphics[width=5.6cm]{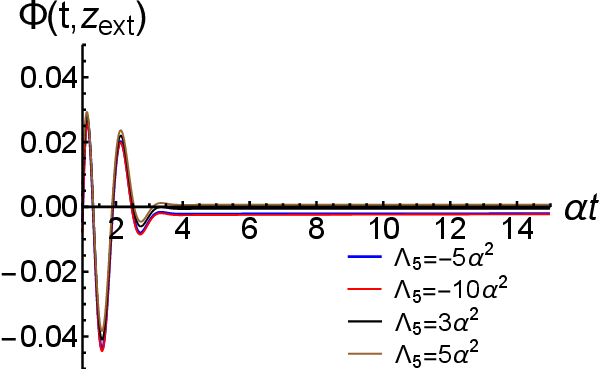}}
    \subfigure[~$|\Phi(t,z_{ext})|$]  {\label{thinlogPhiz1}
    \includegraphics[width=5.6cm]{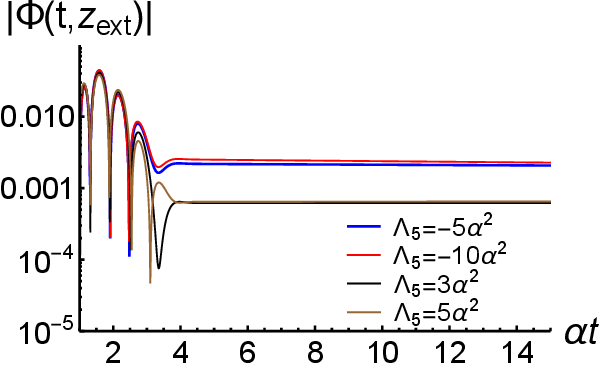}}
    \end{center}
    \caption{Time evolution of the initial wave packet at the extraction point $\alpha z_{ext}=1$ for $p/\alpha=5$. 
    The right figure shows the evolution waveforms on a logarithmic scale.}
    \label{Figure-8}
\end{figure*}

First, we discuss the thin brane case.
The specific forms of 
Eq.~\eqref{psi-equation-re} are as follows:
\begin{equation}
    \left\{
    \begin{aligned}
        &\left[\partial_{\bar{t}}^{2} - \partial_{\bar{z}}^{2} + e^{-2 \bar{t}} \frac{p^2}{\alpha^2}  - 15 \left(\frac{\sqrt{\frac{6\alpha^2}{\Lambda_+}} K_+}{ e^{\bar{z}}  + \frac{6\alpha^2}{\Lambda_+} K_+^2  e^{-\bar{z}} } \right)^2  \right]\Psi(\bar{t},\bar{z})=0 , \quad \bar{z}>0, \\
        &\left[\partial_{\bar{t}}^{2} - \partial_{\bar{z}}^{2} + e^{-2 \bar{t}} \frac{p^2}{\alpha^2} - 15 \left(\frac{\sqrt{\frac{6\alpha^2}{\Lambda_-}} K_-}{ e^{-\bar{z}}  + \frac{6\alpha^2}{\Lambda_-} K_-^2  e^{\bar{z}} } \right)^2  \right]\Psi(\bar{t},\bar{z})=0 , \quad \bar{z}<0,
    \end{aligned}
    \right.
    \label{wave-equation-thin-origin}
\end{equation}
Then, we choose a Gaussian function as the initial wave packet, 
$\Psi_{\text{ini}}(0,\bar{z}) \sim e^{-\frac{\kappa^2 \bar{z}^2}{2}}$, 
and the boundary conditions are as follows:
\begin{equation}
    \left\{
    \begin{aligned}
        &\partial_{\bar{t}} \Psi = -\partial_{\bar{z}} \Psi , \quad &&\bar{z} \rightarrow \infty, \\
        &\partial_{\bar{z}} \Psi |_{\bar{z}=0^+} - \partial_{\bar{z}} \Psi |_{\bar{z}=0^-} =  -\frac{\sigma}{\alpha} \Psi(\bar{t},\bar{z} = 0) ,  \quad &&  \\
        &\partial_{\bar{t}} \Psi = \partial_{\bar{z}} \Psi , \quad &&\bar{z} \rightarrow -\infty.
    \end{aligned}
    \right.
    \label{boundary-condition-thin-time-evolution}
\end{equation}
Our purpose is to solve $\Phi(t,z)$, 
which is related to $\Psi(t,z)$ with $\Phi(t,z) \sim e^{-\frac{3}{2}\alpha t} \Psi(t,z)$
specified in Eq.~\eqref{Phi-assumption}.
The evolution waveforms of $\Phi(t,z)$ are shown in Fig.~\ref{Figure-8}.

As shown in Fig.~\ref{Figure-8}, 
the initial stage of the waveform evolution exhibits oscillatory behavior, 
primarily governed by the initial wave packet.  
Subsequently, the waveform amplitude gradually remains unchanged;
fitting results indicate a frequency of $\omega^{\text{tot}}_{0}/\alpha = 1.34579\times 10^{-6} + 0.00109 i$, 
which corresponds to the zero mode $\omega^{\text{tot}}_{0} = 0$.
Moreover, regardless of the choice of the bulk cosmological constant, 
the late-time evolution remains consistent,
ultimately dominated by the zero mode. 

%%%%%%%%%%%%%%%%%%%%%%%%%%%%%%%%%%%%%%%%%%%%%%%%%%%%%%%%%%%%%%%%%%%%%%%%%%%%%%%%%%%%%%%%%%%%%%%%%%%%%%%%%%%%%%%%%%%%%%%%%%%%%
\subsubsection{Thick branes} \label{Time domain evolution of dS thick branes}

\begin{figure*}[htb]
    \begin{center}
    \subfigure[~$\Phi(t,z_{ext})$]  {\label{thickPhiorigin}
    \includegraphics[width=5.6cm]{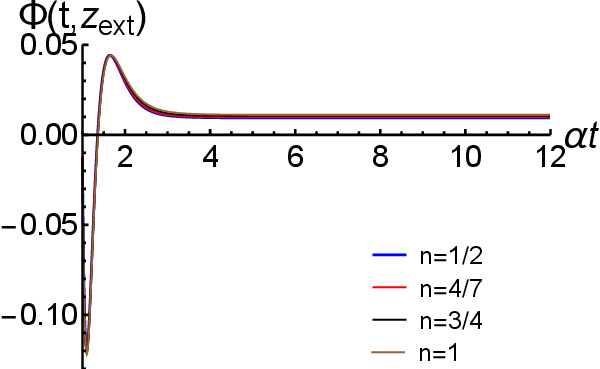}}
    \subfigure[~$|\Phi(t,z_{ext})|$]  {\label{thicklogPhiorigin}
    \includegraphics[width=5.6cm]{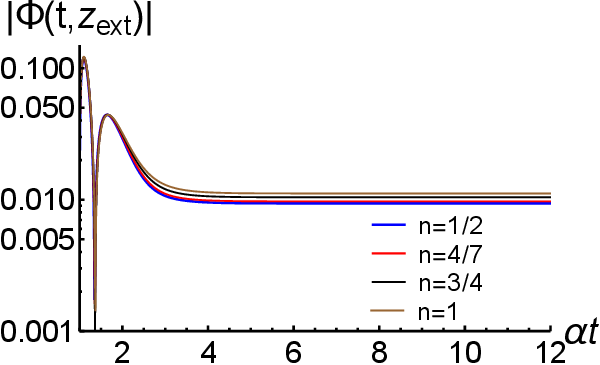}}
    \end{center}
    \caption{Time evolution of the initial wave packet of $\Phi(t,z)$
    at the extraction point $\alpha z_{ext}=1$ for $p/\alpha=10$. 
    The right figure shows the evolution waveforms on a logarithmic scale.}
    \label{Figure-12}
\end{figure*}

\begin{figure*}[htb]
    \begin{center}
    \subfigure[~$\hat{\Phi}(t,z_{ext})$]  {\label{thickPhidual}
    \includegraphics[width=5.6cm]{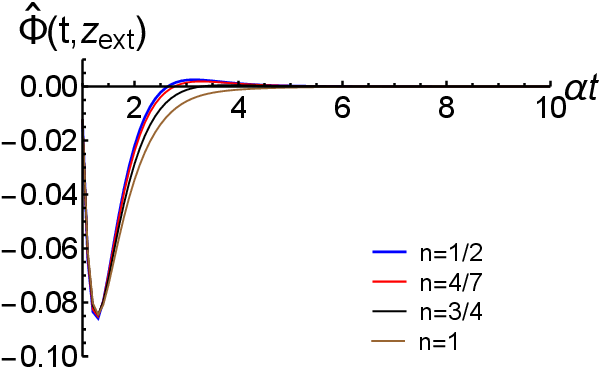}}
    \subfigure[~$|\hat{\Phi}(t,z_{ext})|$]  {\label{thicklogPhidual}
    \includegraphics[width=5.6cm]{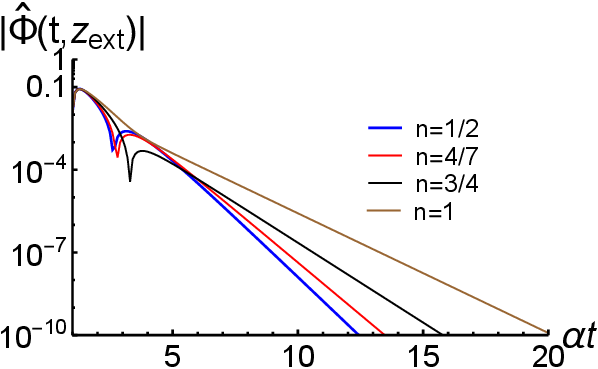}}
    \end{center}
    \caption{Time evolution of the initial wave packet of $\hat{\Phi}(t,z)$
    at the extraction point $\alpha z_{ext}=1$ for $p/\alpha=5$. 
    The right figure shows the evolution waveforms on a logarithmic scale.}
    \label{Figure-13}
\end{figure*}

Since the junction condition is unnecessary in the thick brane case, 
we can simplify the process by using the light-cone coordinates 
$d\bar{u}=d\bar{t}-d\bar{z}$ and $d\bar{v}=d\bar{t}+d\bar{z}$. 
The specific forms of Eqs.~\eqref{psi-equation-re} and~\eqref{psi-equation-re-dual} can be rewritten as follows: 
\begin{align}
    \label{uv-wave-equation-origin}
    \left(4 \frac{\partial^{2}}{\partial_{\bar{u}}\partial_{\bar{v}}} + e^{-2 \bar{t}} \frac{p^2}{\alpha^2} - \frac{3n(3n+2)}{4 n^2}  \frac{1}{\cosh^2(\bar{z}/n)}  \right)\Psi(\bar{u},\bar{v}) &=0,   \\
    \label{uv-wave-equation-dual}
    \left(4 \frac{\partial^{2}}{\partial_{\bar{u}}\partial_{\bar{v}}} + e^{-2 \bar{t}} \frac{p^2}{\alpha^2} + \frac{3n(2-3n)}{4 n^2} \frac{1}{\cosh^2(\bar{z}/n)}  \right)\hat{\Psi}(\bar{u},\bar{v}) &=0.   
\end{align}
Then, we choose the initial incident wave packet as follows: 
\begin{align}
    \label{incident-wave-packet-even}
    &\Psi_{\text{ini}}(\bar{u},0)=\hat{\Psi}_{\text{ini}}(\bar{u},0)=\cos\left(\kappa \bar{u}\right) e^{-\frac{\kappa^2 \bar{u}^2}{2}}, \\
    &\Psi_{\text{ini}}(0,\bar{v})=\hat{\Psi}_{\text{ini}}(0,\bar{v})=\cos\left(\kappa \bar{v}\right) e^{-\frac{\kappa^2 \bar{v}^2}{2}}.
\end{align}
With the above initial functions, according to $\Phi(t,z) \sim e^{-\frac{3}{2}\alpha t} \Psi(t,z)$ and $\hat{\Phi}(t,z) \sim e^{-\frac{3}{2}\alpha t} \hat{\Psi}(t,z)$, 
we plot the evolution waveforms of $\Phi(t,z)$ in Figs.~\ref{Figure-12} and~\ref{Figure-13}. 
\begin{center}
    \begin{table}[!htb]
    \renewcommand\arraystretch{1.0}
    \begin{tabular}{|c|c|c|}
    \hline
    ~~$n$~~  &~~~~Analytical method~~~~ &~~Time-domain evolution method~~\\
    \hline
        &~$\text{Re}(\omega^{\text{tot}}_{1}/\alpha) $~~~~$\text{Im}(\omega^{\text{tot}}_{1}/\alpha) $~&~$\text{Re}(\omega^{\text{tot}}_{1}/\alpha) $~~~~$\text{Im}(\omega^{\text{tot}}_{1}/\alpha) $~ \\
    \hline
    ~~$\tfrac{1}{2}$~~ &   ~~0 ~~~~-2.000000 & 8.76337$\times 10^{-4}$~~~~-1.999995 \\
    \hline
    ~~$\tfrac{4}{7}$~~ &   ~~0 ~~~~-1.750000& 4.54248$\times 10^{-5}$~~~~-1.750028 \\
    \hline
    ~~$\tfrac{3}{4}$~~ &   ~~0 ~~~~-1.333333& 3.54285$\times 10^{-6}$~~~~-1.333338 \\
    \hline
    ~~$1$~~ &   ~~0 ~~~~-1.000000& 1.31767$\times 10^{-7}$~~~~-0.999999 \\
    \hline
    \end{tabular}\\
    \caption{The first frequency $\omega^{\text{tot}}_{1}$ using the analytical method and the time-domain evolution method.}
    \label{Table-1}
    \end{table}
\end{center}

From Fig.~\ref{Figure-12}, regardless of the choice of the parameter $n$, 
the evolution results remain consistent. 
By fitting the late-time data of the waveform, 
the resulting QNF is $\omega^{\text{tot}}_{0}/\alpha=4.05955\times 10^{-7} + 0.000001 i$, 
which corresponds to the zero mode. 
In contrast, Fig.~\ref{Figure-13} shows that the evolution results are influenced by the parameter $n$.
As indicated in Table~\ref{Table-1}, 
the fitted late-time waveform aligns closely with the analytical results.
The decay of the waveform in the late stage is determined 
by the first QNF $\omega^{\text{tot}}_{1}/\alpha$,
which aligns with the analytical results shown in Fig.~\ref{Figure-7}. 
Moreover, by combining Table~\ref{Table-1} and Fig.~\ref{Figure-13}, 
it can be observed that at $\alpha z_{ext}=1$,
the wave amplitudes initially increase and then gradually decay. 
The initial stage is due to the propagation of the wave packet, 
while the subsequent decay is dominated by the first QNM. 

Based on the results from Figs.~\ref{Figure-12} and~\ref{Figure-13} and Table~\ref{Table-1}, 
although $\hat{\Psi}$ is not the actual waveform of the perturbation field, 
it can be used to fit the first QNM of the actual waveform $\Phi(t,z)$. 
This not only validates that the ``duality relation'' 
between Eqs.~\eqref{psi-equation-re} and~\eqref{psi-equation-re-dual} is credible to some extent, 
but also aligns with our analytical result prediction.

\begin{figure*}[htb]
    \begin{center}
    \subfigure[~$\Phi(t,z_{ext})$]  {\label{pPhiorgin}
    \includegraphics[width=5.6cm]{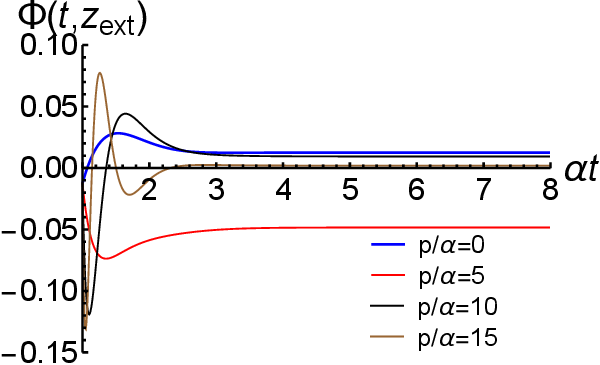}}
    \subfigure[~$|\Phi(t,z_{ext})|$]  {\label{plogPhiorgin}
    \includegraphics[width=5.6cm]{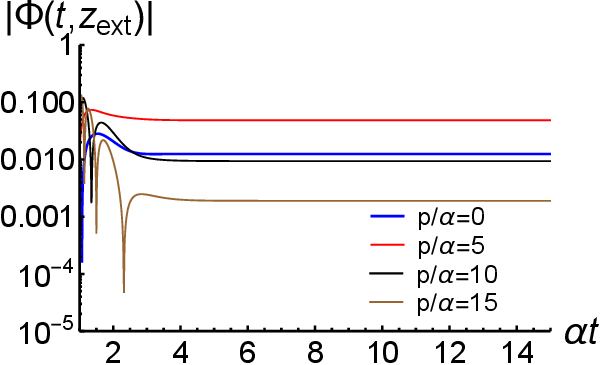}}\\
    \subfigure[~$\hat{\Phi}(t,z_{ext})$]  {\label{pPhidual}
    \includegraphics[width=5.6cm]{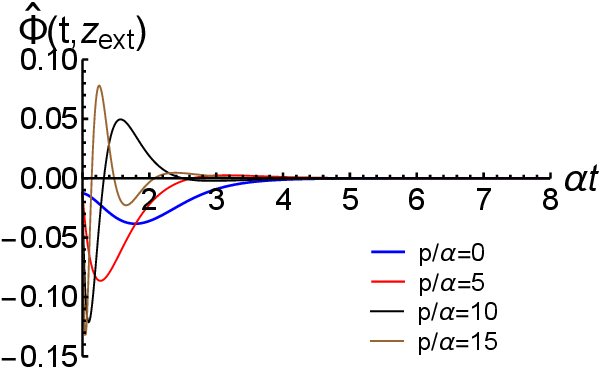}}
    \subfigure[~$|\hat{\Phi}(t,z_{ext})|$]  {\label{plogPhidual}
    \includegraphics[width=5.6cm]{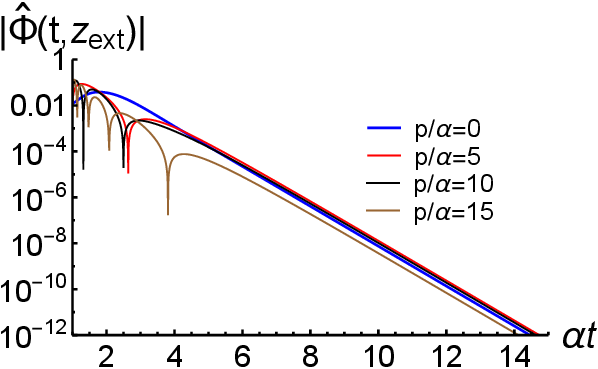}}
    \end{center}
    \caption{Time evolution of the initial wave packet at the extraction point $\alpha z_{ext}=1$
    for the cases of $p/\alpha = 0,5,10,15$ for $n=1/2$.
    The right panel figures show the evolution waveforms on a logarithmic scale.}
    \label{Figure-14}
\end{figure*}

Next, we discuss the effect of the three-dimensional spatial momentum $p$ on the waveform evolution, 
focusing on the numerical evolution of Eqs.~\eqref{psi-equation-re} and~\eqref{psi-equation-re-dual} with $n=\tfrac{1}{2}$, 
as illustrated in Fig.~\ref{Figure-14}.
It can be seen that as $p$ increases, 
the oscillatory decay time in the initial stage of the waveform evolution becomes longer. 
But it quickly transits to a purely stable (or decay) form dominated by the zero mode (or the first QNM).
This indicates that, regardless of the magnitude of the three-dimensional momentum $p$,
the late-time behavior is always consistently governed by the zero mode and the QNMs. 
And $p$ affects only the early-stage waveform frequencies, amplitudes, and decay rates. 

In summary, based on the simulated evolution, 
the actual wave frequency obtained from the evolution equation~\eqref{psi-equation-re} 
is expected to be time-dependent. 
When $t \gg t_0$, the late stage of the waveform evolution is dominated by the zero mode and the QNMs. 
And it can be shown that all modes are stable, that is $\text{Im}(\omega^{\text{tot}})\leq 0$.

To gain a better understanding of the phenomenology of these QNMs, 
we consider a wave packet of gravitational radiation on the dS brane. 
Following Ref.~\cite{Seahra:2005iq},
we propose that this wave packet consists of a superposition of discrete pulses corresponding to the zero mode and all QNMs, 
expressed in the following form:
\begin{equation}
    \label{wave-packet-hij}
    \delta h_{ij}\sim\epsilon_{ij}\int dp \; \vartheta(p) \sum_{j=0} c_j \; e^{-i \omega^{\text{tot}}_{j} t -ipx } ,
\end{equation}
where $\vartheta(p)$ is the momentum space profile of the wave packet 
and the expansion coefficients $c_j$ are determined by the initial extra-dimensional pulse profile. 
In Eq.~\eqref{wave-packet-hij}, the first term $j=0$ represents the contribution from the zero mode, 
corresponding to the gravitational wave in the four-dimensional cosmological background. 
The subsequent terms are dominated by the QNMs $\omega^{\text{tot}}_{j}$ ($j>0$), 
which have vanishing real parts and negative imaginary parts.
These terms correspond to the massive gravitons and 
will lead to a purely decaying behavior of the waveforms, as shown in Fig.~\ref{Figure-14}.
From this, we can see that the part associated with the QNMs contains information about the Hubble constant $\alpha$. 
Given the current observational data on the Hubble constant, around $10^{-10}\;\text{yr}^{-1}$~\cite{Riess:2016jrr,Planck:2018vyg}, 
the decay lifetime of the actual waveform is extremely long, 
approximately a billion years, which is about roughly a few percent of the Universe's age.
Therefore, we can speculate that the quasinormal signals from the extra dimension
may be hidden within the background of stochastic gravitational waves.
Through detecting such signals, we can not only probe the structure of the extra dimension 
and validate the braneworld hypothesis, 
but also infer the Hubble constant with the actual observational data. 
This provides us a new measurement method that is worth further exploration and verification.

%%%%%%%%%%%%%%%%%%%%%%%%%%%%%%%%%%%%%%%%%%%%%%%%%%%%%%%%%%%%
\section{Conclusion} \label{conclusion}

In this paper, we investigated the quasinormal ringing behavior of dS branes.
Specifically, we analyzed the transverse-traceless gravitational perturbations in both the thin and thick brane scenarios. 

For the thin dS brane, we found that the bulk cosmological constant $\Lambda_5$ directly affects the shapes of the effective potentials, 
as shown in Fig.~\ref{Figure-5}. Additionally, there is a negative delta function at the location of the brane ($z=0$). 
By analytically solving the Schrödinger-like equation governing the gravitational perturbations on the brane, 
we further discovered that the QNMs are purely imaginary,
which is given by Eq.~\eqref{QNMs-result-total} with $n=1$, 
regardless of the cosmological constants on either side of the brane, and these modes depend solely on the Hubble constant $\alpha$.
For the thick dS brane, the shape of the effective potential is affected by the parameter $n$,
as shown in Fig.~\ref{Figure-6}. 
By utilizing the analytical and continued fraction methods 
to solve Eq.~\eqref{PT-equation-re-thick},
we found that the QNMs~\eqref{QNMs-result-total} remain purely imaginary. 
These results are dependent on both the Hubble constant $\alpha$ and the parameter $n$ that represents the internal structure of the brane.
Therefore, we could conclude that the dS branes possess purely imaginary QNMs, 
and these modes depend on the internal structure of the brane as well as the Hubble constant.

The QNMs of the dS branes exhibit a particular uniqueness, which is also present in other black holes 
and braneworld scenarios~\cite{Jia:2024pdk,Konoplya:2008yy,Konoplya:2008au,Cuyubamba:2016cug,Konoplya:2022xid,Sarkar:2023rhp,Bohra:2023vls}. 
We speculate that this phenomenon may be related to the RG flow~\cite{Bazeia:2006ef}. 
On the other hand, based on our findings in the simulated evolution of the gravitational wave packets on the brane, 
as shown in Fig.~\ref{Figure-14} and Eq.~\eqref{wave-packet-hij}, 
all QNMs, except for the zero mode, exhibit a purely decaying behavior, 
which contain the Hubble constant information. 
If these QNMs are detected, they will provide new insights into the measurement of the Hubble constant.

%%%%%%%%%%%%%%%%%%%%%%%%%%%%%%%%%%%%%%%%%%%%%%%%%%%%%%%%%%%%%%%%%%%%%%%%%%%%%%%%%%%%%%%%%%%%%%%%%%%%%%%%%%%%%%%%%%%%%%%%%%%%
\section*{Acknowledgments}
We would like to thank Wen-Yi Zhou for very useful discussions. 
This work was supported by 
the National Key Research and Development Program of China (Grant No. 2021YFC2203003), 
the National Natural Science Foundation of China (Grants No. 12475056, No. 12247101, No. 12205129, No. 12347111, and No. 12405055), 
the China Postdoctoral Science Foundation (Grants No. 2023M741148), 
the Fundamental Research Funds for the Central Universities (Grants No. lzujbky-2025-it05), 
the 111 Project (Grant No. B20063), 
the Postdoctoral Fellowship Program of CPSF (Grant No. GZC20240458), 
Gansu Province's Top Leading Talent Support Plan.
\par

Data Availability Statement. This article has no associated data or the data will not be deposited.
\par

Code Availability Statement. This article has no associated code or the code will not be deposited.
\par

\end{document}